\documentclass[11pt,a4paper]{article}

\usepackage{microtype}

\usepackage{bbm} 
\usepackage{mathrsfs} 

\usepackage{todonotes}

\usepackage{etoolbox}

\usepackage{mathtools} 

\usepackage{listings}

\usepackage{nccmath}

\usepackage{jheppub}
\usepackage{pdfsync}

\usepackage{epsfig,epsf}
\usepackage{amsmath}
\usepackage{amsthm}
\usepackage{amsfonts}
\usepackage{amssymb}
\usepackage{dsfont}
\usepackage{marvosym}
\usepackage{multirow}

\RequirePackage{graphicx}
\RequirePackage{flushend}
\RequirePackage[numbers,sort&compress]{natbib}

\usepackage{slashed}
\usepackage{epstopdf}
\usepackage[active]{srcltx}

\def\II{\hbox{{1}\kern-.25em\hbox{l}}}

\newcommand \vev [1] {\langle{#1}\rangle}
\newcommand \VEV [1] {\left\langle{#1}\right\rangle}

\newcounter{VBQ}

\title{Integrability in heavy quark effective theory}

\author[a]{Vladimir M. Braun,}
\author[a]{Yao Ji\,}
\author[b,a,c]{and Alexander N.  Manashov\,}

\subheader{
\begin{flushright}
{\small
DESY 18-056
}
\end{flushright}
}

\affiliation[a]{
   Institut f\"ur Theoretische Physik, Universit\"at
   Regensburg,  D-93040 Regensburg, Germany}
\affiliation[b]{
   Institut f\"ur Theoretische Physik, Universit\"at Hamburg,
   D-22761 Hamburg, Germany}

\affiliation[c]{ St.Petersburg Department of Steklov
Mathematical Institute,
191023 St.Petersburg, Russia
 }
\emailAdd{vladimir.braun@physik.ur.de}

\emailAdd{yao.ji@ur.de}

\emailAdd{alexander.manashov@desy.de}

\abstract{
It was found that renormalization group equations in the heavy-quark effective theory (HQET) 
for the operators involving one effective
heavy quark and light degrees of freedom are completely integrable in some cases and are related to
spin chain models with the Hamiltonian commuting with the nondiagonal entry $C(u)$ of the monodromy matrix.
In this work we provide a more complete mathematical treatment of such spin chains in the QISM framework.
We also discuss the relation of integrable models that appear in the HQET context with the large-spin limit
of integrable models in QCD with light quarks.
We find that the conserved charges and the ``ground state'' wave functions in HQET models can be obtained from
the light-quark counterparts in a certain scaling limit.
       }


\begin{document}

\maketitle

\newpage
\section{Introduction}

The notion of effective field theories (EFT) is central to modern particle physics both at the conceptual level and
as a calculational tool. In particular the large range of quark masses in nature invites the EFT construction.
If the quarks are very light as compared to the typical scales, their masses effectively become irrelevant and can be treated
as a perturbation; the theory of strong interactions in this limit acquires additional --- chiral --- symmetry and can be
matched to an effective low-energy theory described by chiral Lagrangian. Similarly, if the quarks are very heavy,
their masses again become irrelevant. Heavy quarks in loops decouple whereas heavy quarks in initial and final states
move along their classical trajectories and can be thought of as sources of an external Coulomb field.
The corresponding EFT --- the Heavy Quark Effective Theory (HQET)~\cite{Neubert:1993mb} --- is well established and
contributed significantly to the studies of flavor sector of the Standard Model.

Gauge theories can have ``hidden'' symmetries that are not seen at the Lagrangian level.
In particular, it turns out that the renormalization group equations (RGEs) in QCD are integrable for several important cases
to one loop accuracy in the multi-color limit~\cite{Lipatov:1993yb,Faddeev:1994zg,Braun:1998id,Braun:1999te,Belitsky:1999bf}.
This property allows one to apply a powerful mathematical apparatus -- Quantum Inverse Scattering Method
(QISM)~\cite{Faddeev:1979gh,Takhtajan:1979iv,Kulish:1981gi,Sklyanin:1995bm} -- to study properties of these equations and their
solutions in great detail, see e.g.~\cite{Braun:1999te,Braun:2000av,DeVega:2001pu,Derkachov:2001yn,Braun:2009vc} for several
concrete applications. Integrability was also discovered in the ${\cal N}=4$ supersymmetric Yang-Mills theory~\cite{Minahan:2002ve,Beisert:2010jr}
 and is much more powerful in this case.
A comprehensive review of integrability in ${\cal N}=4$ supersymmetric Yang-Mills theory and its connection with the AdS/CFT correspondence
can be found in~\cite{Beisert:2010jr}, together with further references.

It is natural to expect that an EFT describing a certain sector of the underlying theory retains some of the
symmetries. Indeed, it was found \cite{Braun:2014npa,Braun:2015pha,Braun:2017liq} that RGEs in HQET for the operators involving one effective
heavy quark and light degrees of freedom are integrable under similar conditions as in QCD with light quarks and are related to unconventional
integrable models with the Hamiltonian commuting with the nondiagonal entry $C(u)$ of the monodromy matrix.
Analogous unconventional integrable models have appeared recently in the studies of high-energy scattering amplitudes
in the ${\cal N}=4$ supersymmetric Yang-Mills theory ~\cite{Lipatov:2009nt,Basso:2010in,Bartels:2011nz,Belitsky:2011nn}.

In Refs.~\cite{Braun:2014npa,Braun:2015pha,Braun:2017liq} only the results of immediate relevance for the phenomenology of $B$-meson
weak decays were presented without derivation.
The aim of the present paper is twofold. First, we
provide a more complete mathematical treatment of the RG equations in HQET in the framework of QISM approach.
The relevant spin chain models are identified and solved for one heavy and arbitrary number of light degrees of freedom.

Second, we discuss the relation of integrable models that appear in the HQET context with the large-spin limit
of integrable models in QCD with light quarks~\cite{Braun:1998id,Braun:1999te,Belitsky:1999bf}.
We find that the conserved charges in HQET models can be obtained from the light-quark counterparts by a simple rescaling
procedure, and also the ``ground state'' wave functions are related.

\section{General remarks}

\subsection{Renormalization Group Equations in HQET}\label{sect:HQET}

For our discussion the two-component spinor formalism is the most convenient.
We write the Dirac spinor as
\begin{align}
 q=\begin{pmatrix}
\psi_\alpha\\\bar \chi^{\dot\beta}\end{pmatrix}\,, \qquad\qquad \bar q=(\chi^\beta,\bar\psi_{\dot\alpha})
\end{align}
and decompose the gluon field strength in terms of  chiral and antichiral symmetric tensors
 $f_{\alpha\beta}$ and $\bar f_{\dot\alpha\dot\beta}$,
\begin{align}
F_{\alpha\beta,\dot\alpha\dot\beta} =\sigma^\mu_{\alpha\dot\alpha} \sigma^\nu_{\beta\dot\beta} F_{\mu\nu}=
2\left(\epsilon_{\dot\alpha\dot\beta} f_{\alpha\beta}-
\epsilon_{\alpha\beta} \bar f_{\dot\alpha\dot\beta}\right),
\end{align}
which belong to $(1,0)$ and $(0,1)$ representations of the Lorenz group, respectively.

Twist decomposition is usually done by a projection on a pair of auxiliary light-like vectors
$n^2=0$, $\bar n^2=0$ which can be represented by a product of auxiliary spinors
\begin{align}
n_{\alpha\dot\alpha}=n_\mu\sigma^\mu_{\alpha\dot\alpha}=\lambda_\alpha\bar\lambda_{\dot\alpha}\,,
&& \bar n_{\alpha\dot\alpha}=\bar n_\mu\sigma^\mu_{\alpha\dot\alpha}=\mu_\alpha\bar\mu_{\dot\alpha}
\end{align}
where $\bar\lambda = \lambda^\dagger$, $\bar\mu = \mu^\dagger$.
The ``+'' and ``--'' fields are defined as
\begin{align}
\chi_+=\lambda^\alpha \psi_\alpha, && \bar\psi_+=\bar\lambda^{\dot \alpha} \psi_{\dot\alpha},
&& f_{++}=\lambda^\alpha\lambda^\beta f_{\alpha\beta}, && f_{+-}=\lambda^\alpha\mu^\beta f_{\alpha\beta}\,,
&&
\bar f_{++}=\bar\lambda^{\dot \alpha}\bar\lambda^{\dot\beta} \bar f_{\dot\alpha\dot\beta}
\end{align}
etc.
The effective heavy quark field of HQET $h_v$ can be represented by a Wilson line in a timelike direction
$v = (1/2) (n+\bar n)$, $v^2=1$
with an attached free Dirac spinor so that~\cite{Korchemsky:1991zp}:
\begin{equation}
       \langle 0| h_v(0) |h,v\rangle = [0,-v\infty] = {\rm Pexp}\left[ig\int_{-\infty}^0\!d\alpha\,v_\mu A^\mu(\alpha v)\right]\,.
\label{hv}
\end{equation}
The equation of motion (EOM) $\slashed{v}h_v = h_v $ implies for the two-component spinors
\begin{align}
    h_{+} = - \bar h_-\,, \qquad h_- =  \bar h_+\,.
\label{EOMheavy}
\end{align}

In this work we will be dealing with renormalization of gauge-invariant operators built of a heavy quark and light quark/gluon
fields at lightlike separations (``light-ray operators''). The simplest operator in question is
\begin{align}
   {O}_+(z,\mu) &=  \bar \psi_+(zn) [zn,0] h_+(0)\,, &&
 {}[zn,0] = {\rm Pexp}\left[ig\int_0^1\!d\alpha\,n_\mu A^\mu(\alpha z n)\right].
\end{align}
Thanks to \eqref{hv}
this operator can be viewed as a single light antiquark attached to the Wilson line with a cusp containing  one lightlike and one timelike segment.
Its matrix element between vacuum and HQET meson state defines what is called a leading twist
heavy-meson (e.g. $B$-meson in static limit) distribution amplitude (DA) in position space
\begin{align}
  i \langle 0| {O}_+(z,\mu) |B(v)\rangle \sim  \Phi_+(z;\mu)\,.
\end{align}
The DA $\Phi_+(z;\mu)$ is an analytic function of the light-cone separation $z$ in the lower half of the complex plane.
Its scale dependence is driven by the RGE for the operator ${O}_+(z,\mu)$, which has the form
\begin{align}
\label{RGE1}
 \left(\mu\frac{\partial}{\partial\mu}+\beta(g) \frac{\partial}{\partial g}+\frac{\alpha_s C_F }{\pi}\mathcal{H}_{qh}\right)
{O}_+(z,\mu)=0\,,
\end{align}
where the evolution kernel $\mathcal{H}_{qh}$ (the heavy-light ``Hamiltonian'') is an integral operator~\cite{Lange:2003ff,Braun:2003wx,Knodlseder:2011gc}
\begin{align}
\label{HLN}
{} [\mathcal{H}_{qh}f](z)= \int_0^1\frac{d\alpha}{\alpha}\Big(f(z)-\bar\alpha f(\bar\alpha z)\Big)
+\ln(i\mu z)\,f(z)-(\sigma_h+\sigma_q)\,f(z)\,, \qquad \bar\alpha \equiv 1-\alpha\,.
\end{align}
Here $\sigma_h = 1/2$ and $\sigma_q=3/4$ are the heavy and light quark anomalous dimensions, respectively.
In what follows we imply using dimensional regularization with minimal subtraction ($\overline{\text{MS}}$-scheme).

It was noticed~\cite{Braun:2014owa} that this operator can be written in a simpler form in terms of the generator of special
conformal transformations
\begin{align}
  \mathcal{H}_{qh} = \ln (i\mu S^+_q) +\gamma_E-\sigma_h-\sigma_q\,.
\label{Hqh}
\end{align}
Thus $\mathcal{H}_{qh}$ and $S^+_q =  z^2 \partial_z + 2 z$ share the same eigenfunctions
\begin{align}\label{eigenS+}
iS_+\,Q_s(z) =   s \,Q_s(z)\,,
&&
\mathcal{H}_{qh}\, Q_s(z) & = \Big[ \ln(\mu\,s) +\gamma_E -\sigma_h-\sigma_q \Big]\,Q_s(z)
\end{align}
with
\begin{align}\label{eigenH}
Q_s(z)=-\frac{1}{z^2} e^{is/z}\,,
\end{align}
providing the complete set of solutions for the RGE \eqref{RGE1}.

In the description of heavy baryons and also of higher Fock states in heavy mesons, more complicated
operators arise that involve more than one light degree of freedom, of the type
\begin{align}
   &  \psi_+(z_1n)  \psi_+(z_2n) h_+(0)\,,
\notag\\
   &   \chi_+(z_1n)  \bar f_{++}(z_2n) h_+(0)\,,
\notag\\
   &   \chi_+(z_1n)  f_{+-}(z_2n) h_-(0)\,,
\label{examples}
\end{align}
etc., where we suppress color structure and the gauge links.
The ``Hamiltonians'' appearing in the RGEs for such operators to one loop accuracy have a pairwise
structure, e.g.
\begin{align}
  \mathcal{H}_{qgh} = \mathcal{H}_{qg} + \mathcal{H}_{qh} + \mathcal{H}_{gh}\,,
\label{HQET-H}
\end{align}
where the ``heavy-light'' two-particle evolution kernels have the form similar to \eqref{Hqh} with the generators in the
appropriate representation, and the ``light-light'' ones can be written in terms of the corresponding
quadratic Casimir operators~\cite{Bukhvostov:1985rn}. Explicit expressions can be found, e.g., in~\cite{Braun:2017liq}.

It turns out~\cite{Braun:2014npa,Braun:2015pha,Braun:2017liq} that these more complicated RGEs are completely integrable and
can be solved using QISM techniques. 
In this work we construct and discuss the corresponding spin chain models which differ somewhat from the standard ones and may be interesting
in other applications.

\subsection{Spin chain models}\label{sect:chains}
The RGE kernels in HQET of the type \eqref{HQET-H} can  be identified with  the Hamiltonians of certain spin chain models with
$\mathrm{SL}(2,\mathbb{R})$ symmetry. These models describe   quantum mechanical systems of interacting spins,
$S^{(k)}_\alpha=\{S^{(k)}_+,S^{(k)}_-,S^{(k)}_0\}$, which are the generators of
$\mathrm{SL}(2,\mathbb{R})$. The index  $k=1,\ldots,N$ enumerates the sites of the chain,
where the number $N$ corresponds to the number of light degrees of freedom.
The spin operators on a given site $k$ obey standard commutation relations
\begin{align}
[S^{(k)}_0, S_{\pm}^{(k)}]=\pm S_{\pm}^{(k)}\,, && [S^{(k)}_+, S_{-}^{(k)}]=2 S_{0}^{(k)}
\end{align}
and  commute with each other
for $k\neq k'$. The generators can conveniently be realized as the first order differential operators
\begin{align}\label{generators}
S_+^{(k)}=z_k^2\partial_{z_k} + 2 z_k s_k,
&&
S_0^{(k)}=z_k\partial_{z_k} + {s_k},
&&
S_-^{(k)}=-\partial_{z_k}\,.
\end{align}
The spin $s_k$ labels  a representation of the $\mathrm{SL}(2,\mathbb{R})$ group. The choice of the representation depends on
the problem under consideration. In statistical physics one usually encounters spin chains with finite dimensional
representations, while in QFT one deals with infinite dimensional representations. In the present context we need the
so-called  discrete series representation of
$\mathrm{SL}(2,\mathbb{R})$ group, $D_s^{-}$~\cite{MR3469700}. It is defined on the space of functions analytic in the lower half-plane and
equipped with the scalar product~\cite{MR3469700}
\begin{align}\label{scalar-product}
{\vev{f,g}_s}=\int\limits_{\text{Im}\,z<0}\!\! D_sz \;(f(z))^* g(z)\,,
\end{align}
where
\begin{align}\label{measure}
D_s z= \dfrac{2s-1}\pi (-2\text{Im}\, z)^{2s-2} dx dy\,, \qquad z = x+ iy\,.
\end{align}
This scalar product is invariant with respect to the
symmetry transformations
\begin{align}\label{Tg}
f(z)\mapsto [T(g) f](z)= \frac{1}{(cz+d)^{2s}} f\left(\frac{az+b}{cz+d}\right)\,,
\end{align}
where $g^{-1}=\begin{pmatrix} a & b\, \\ c & d\end{pmatrix}\in \mathrm{SL}(2,\mathbb{R})$.
 The operators~\eqref{generators} are the generators of infinitesimal transformations corresponding to~\eqref{Tg}. They are anti-hermitian
with respect to the scalar product~\eqref{scalar-product}.

The Hilbert space  of the $N$-site  spin chain  is given by the space of functions of $N$ complex variables analytic in the
lower complex half-plane in each variable and equipped with the scalar product
\begin{align}\label{scpN}
{\vev{f,g}_{s_1\ldots s_N}}=\prod_{k=1}^N\int_{\text{Im}z_k<0} D_{s_k} z_k \,(f(z_1,\ldots ,z_N))^* g(z_1,\ldots,z_N)\,.
\end{align}
In the following we will often drop the subscripts $s_1,\ldots,s_N$ if the spins are clear from the context.

\section{Heavy-light spin chain models}\label{sect:HLchains}

\subsection{Closed spin chain}\label{sec:closed}

\subsubsection{Monodromy matrix}

The QISM approach allows one to construct a set of mutually  commuting operators (charges) for  spin chain models as follows.
One defines the so-called Lax operator
\begin{align}\label{Lax}
L_k(u)=u + i\begin{pmatrix}
S^{(k)}_0 & S^{(k)}_-\\
S^{(k)}_+ & -S^{(k)}_0
\end{pmatrix}\,,
\end{align}
where the spectral parameter $u$ is a complex number.   The monodromy matrix is defined as a product of the Lax operators
\begin{align}\label{monodromy-closed}
T_N(u)=L_1(u_1) L_2(u_2)\ldots L_N(u_N) =\begin{pmatrix}
A_N(u) & B_N(u)\\
C_N(u) & D_N(u)
\end{pmatrix},
\end{align}
with $u_k=u+\xi_k$, where the   $\xi_k$ are the so-called impurities and we assume $s_k=s$ unless stated otherwise.
By construction, the entries of the monodromy matrix are polynomials in the spectral parameter
$u$. It can be shown that these operators form commuting families~\cite{Faddeev:1979gh}, i.e.
\begin{align}
[A_N(u),A_N(v)]=0, && [B_N(u),B_N(v)] =0, && [C_N(u),C_N(v)]=0, && [D_N(u),D_N(v)] =0\,.
\end{align}
Since
$[S_\alpha+\sigma_\alpha/2, T_{N}(u)]=0$
where $\sigma_{\alpha}$ are the  Pauli matrices
 and $S_\alpha=\sum_{k=1}^N S_\alpha^{(k)}$ is the operator of total spin, it is easy to show that
\begin{align}
[S_0, A_N]&=[S_0,D_N]=0, && [S_-,B_N]=[S_+,C_N]=0\, ,\notag\\
[S_+,D_N]&=-[S_+,A_N]=C_N\, , && [S_-,A_N]=-[S_-,D_N]=B_N\, .
\label{com2}
\end{align}
In the familiar field-theory applications such as the RG equations for light-ray operators built of light quarks/gluons in
QCD, one deals with the $\mathrm{SL}(2,\mathbb{R})$ invariant systems. For such systems the proper object to consider  is the
transfer matrix,
\begin{align}\label{t-closed}
t_N(u)=A_N(u) + D_N(u), && [t_N(u),t_N(v)]=0,
\end{align}
 which is an invariant operator, $[S_\alpha, t_N(u)]=0$.
It turns out that the transfer matrix of a homogeneous chain without impurities commutes also  with the 
(Hamiltonian) operator
\begin{align}\label{Hik}
\mathbb{H}_N=\sum_{k=1}^N \mathcal{H}_{kk+1}\,,  && \mathcal{H}_{kk+1}=2\big(\psi(J_{kk+1})+{\gamma_E}\big) - 2\sigma_q\,,
\end{align}
where $\psi(x)$ is the polygamma function and $J_{kk+1}$ is the two-particle operator of  conformal spin,
\begin{align}
J_{kk+1}(J_{kk+1}-1)=(\vec{S}^{(k)} + \vec{S}^{(k+1)})^2.
\end{align}
The operator $\mathbb{H}_N$ can be identified with the leading-order evolution kernel for certain RGEs in gauge
theories~\cite{Braun:1998id}. Since $\mathbb{H}_N$ commutes with $t_N(u)$ they share the same set of eigenfunctions 
which can be constructed with the help of QISM~\cite{Faddeev:1979gh,Sklyanin:1995bm}.

\subsubsection{Heavy-light Hamiltonian}

The main new element in the present case is that the evolution kernels in HQET are {\it not} $\mathrm{SL}(2,\mathbb{R})$
invariant. It was shown, however, that at leading order all heavy-light kernels commute with the generator $S_+$ of special
conformal transformations~\cite{Knodlseder:2011gc}. Since $[S_+,C_N]=0$ \eqref{com2} it is natural to expect that the conserved
charges (if there is any hidden symmetry) in the heavy-light sector have to be generated by the $C_N(u)$-entry of the
monodromy matrix. An example of such a system is given by the heavy-light baryon~\cite{Ball:2008fw,Wang:2011uv,Ali:2012pn}
corresponding to a two-site chain and first studied in~\cite{Braun:2014npa}. Motivated by this application, we  consider from
now on a homogeneous ($s_k=s$) closed spin chain without impurities ($\xi_k=0$), but with an arbitrary number of sites $N$.

As the first step, we show that the Hamiltonian
\begin{align}\label{HclosedN}
\mathbb{H}_N= \mathcal{H}_1+\sum_{k=1}^{N-1} \mathcal{H}_{kk+1} +\mathcal{H}_N\,,  
\end{align}
where $\mathcal{H}_{kk+1}$ are defined in~\eqref{Hik} and the boundary Hamiltonians are given by%
\footnote{In this discussion we omit trivial constants corresponding to
the quark wave-function renormalization, cf.~\eqref{Hqh}.}
\begin{align}
\mathcal{H}_1 =  \ln \left (i\mu S^{(1)}_+\right) {+\gamma_E}, && \mathcal{H}_N = \ln \left (i\mu S^{(N)}_+\right) {+\gamma_E}\,,
\end{align}
commutes with the $C_N(u)$-entry of the monodromy matrix.
This statement follows almost immediately from the relation~\cite{Sklyanin:1991ss}
\begin{align}\label{HLL}
[\mathcal{H}_{kk+1}, L_k(u) L_{k+1}(u)]=i \big(L_k(u)-L_{k+1}(u)\big)\,,
\end{align}
which is a consequence of the  defining  $RLL$ relation for the $R$-operator~\cite{Faddeev:1979gh},
\begin{align}
R_{12}(u-v) L_1(u)L_2(v)=L_2(v) L_1(u) R_{12}(u-v),
\label{RLL}
\end{align}
and  its small-$u$ expansion: $R_{12}(u)=P_{12}\Big(1 -iu \,\mathcal{H}_{12}+O(u^2)\Big)$.
{Here $P_{12}$ is the permutation operator, $P_{12}f(z_1,z_2)=f(z_2,z_1)$}.
In addition, $[S_0,\ln S_+]=1$ implies that
\begin{align}\label{SS+}
K_1 \equiv [\ln S^{(1)}_+, L_1(u)]=-i\begin{pmatrix} 1& *\\ 0 & -1
\end{pmatrix},  && %
K_N \equiv [\ln S^{(N)}_+, L_N(u)]=-i\begin{pmatrix} 1& *\\ 0 & -1
\end{pmatrix}.
\end{align}%
Using~\eqref{HLL} and \eqref{SS+} one easily finds
\begin{align}
\Big[\sum_{k=1}^{N-1} \mathcal{H}_{kk+1},T_N(u)\Big] &=-i L_{2}(u)\ldots L_N(u) +iL_1(u)\ldots L_{N-1}(u)
\end{align}
and
\begin{align}
 [\mathcal{H}_1,T_N(u)] = K_1\,L_{2}(u)\ldots L_N(u)\,,
&&
[\mathcal{H}_N,T_N(u)] = \,L_{1}(u)\ldots L_{N-1}(u) K_N.
\end{align}
Adding up all terms one verifies that indeed
\begin{align}
[\mathbb{H}_N,T_N(u)]_{21} = [\mathbb{H}_N,C_N(u)]=0\,.
\end{align}

It is convenient to consider the operator $B_N(u)$ instead of $C_N(u)$ at the intermediate steps,
using the fact that they are unitarily equivalent:
The inversion operator $J$ (which is an unitary operator) intertwines $C_N(u)$ and $B_N(u)$.
The inversion operator is defined as
\begin{align}
[Jf](z)=z^{-2s} f(-1/z)\,, && ||Jf||^2=||f||^2\, ,
\label{inversion}
\end{align}
where $||f||^2=\vev{f,f}_s$.
It intertwines the generators,
$JS_0=-S_0J$,~ $J S_\pm = -S_{\mp} J$, and as a consequence the following relation for the monodromy matrix holds:
\begin{align}
J T_N(u) J^{-1}=  \sigma_2 T_N(u) \sigma_2,
\end{align}
where $\sigma_2$ is the Pauli matrix. Comparing the off-diagonal entries in this relation one gets
\begin{align}
JC_N(u) = -B_N(u) J\,.
\end{align}
Thus $C_N(u)$ and $B_N(u)$ are indeed unitarily equivalent and their eigenfunctions are related to each other by inversion.
The Hamiltonian~\eqref{HclosedN} transforms under inversion into
\begin{align}\label{HamB}
\widetilde{\mathbb{H}}_N{\equiv}J\mathbb{H}_N J^{-1} = \widetilde{\mathcal{H}}_1+\sum_{k=1}^{N-1} \mathcal{H}_{kk+1}
                      +\widetilde{\mathcal{H}}_N\,,
\end{align}
where $\widetilde{\mathcal{H}}_{1(N)}= \ln \big(-i\mu S_-^{(1(N))}\big) +\gamma_E$.

Eigenfunctions of the operator $B_N(u)$ provide the basis for Sklyanin's representation of separated variables.
They have been constructed explicitly in~\cite{Derkachov:2002tf} and are given by the product of layer operators
acting on the exponential function. For the homogeneous chain considered here
\begin{align}\label{PsiB}
\Psi_{\{p,\vec{x}\}}(z_1,\ldots, z_N)= 
\Lambda_N(x_1)\Lambda_{N-1}(x_2)\ldots \Lambda_{2}(x_{N-1}) e^{-ip z}\,,
\end{align}
where
\begin{align}\label{pEigen}
   \vec{x} &= \{x_1,\ldots,x_{N-1}\}\,, \qquad  x_k \in \mathbb{R}\,.
\end{align}
The ``momentum'' $p\in \mathbb{R}_+$ is an eigenvalue of the generator of translations
\begin{align}
  \big( S_-^{(1)} + \ldots + S_-^{(N)} \Big) \Psi_{\{p,\vec{x}\}}(z_1,\ldots, z_N) = i p\, \Psi_{\{p,\vec{x}\}}(z_1,\ldots, z_N)\, .
\end{align}
%
\begin{figure}[t]
\centerline{\includegraphics[width=0.50\textwidth]{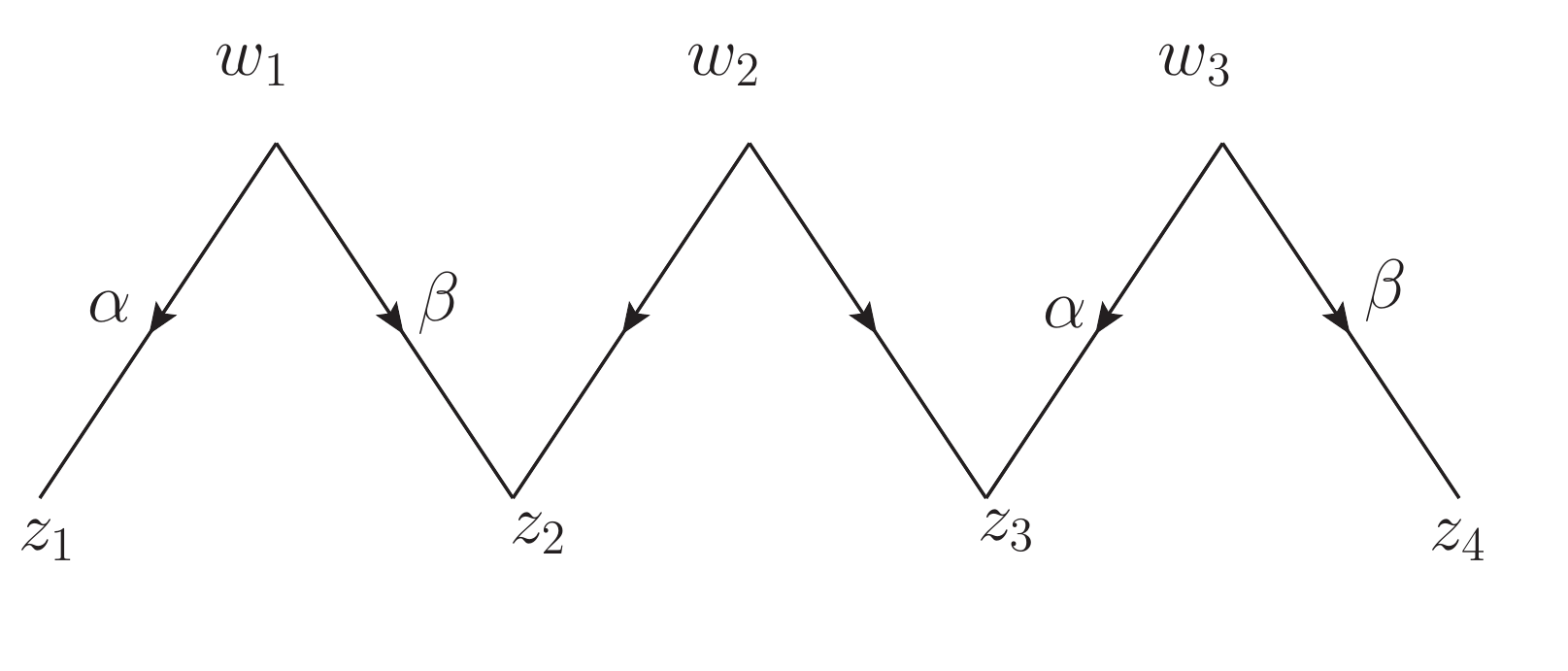}}
\caption{Diagrammatic representation of the layer operator $\Lambda_{N=4}(x)$ for a closed spin chain.
}
\label{fig:Lambda_closed}
\end{figure}
%
The layer operator $\Lambda_M(x)$ 
maps a function of $M-1$ variables into a function of $M$ variables
and is defined as follows:
\begin{align}\label{closedLambda}
[\Lambda_M(x) f](z_1,\ldots,z_M)=\left(\prod_{k=1}^{M-1} \int D_s w_k\right) \Lambda_M(z_1,\ldots,z_M| w_1,\ldots w_{M-1})
\,\, f(w_1,\ldots,w_{M-1})\,.
\end{align}
Here
\begin{align}
\Lambda_M(z_1,\ldots,z_M| w_1,\ldots w_{M-1}) = \prod_{k=1}^{M-1}  D_{s-ix}(z_k,w_k) D_{s+ix}(z_{k+1},w_k)\,,
\label{closedL} \mathbb{}\end{align}
where the function $D_\alpha(z,w)$ ("propagator") is given by the following expression
\begin{align}
D_\alpha(z,w)=\left(\frac {-i}{z-\bar w}\right)^{\alpha}\,.
\end{align}
Here and below $\bar w \equiv w^*$.
The layer operator $\Lambda_M$  can be visualized as the diagram shown in Fig.~\ref{fig:Lambda_closed} where a directed line
from
$w$ to $z$ with an index $\alpha$ stands for the ``propagator''
 $D_\alpha(z,w)$
and the $\alpha$ and $\beta$ indices take the values $s-ix$ and  $s+ix$, respectively.
A summary of the properties of the layer operators can be found in~\cite{Derkachov:2016dhc} (see also Sec.~\ref{sec:Eigenfunctions}).
We mention here that ${\Psi_{\{p,x_1,\ldots,x_{N-1}\}}}$ are symmetric functions of separated variables $x_1,\ldots,x_{N-1}$ and they are
orthogonal to each other with respect to the scalar product~\eqref{scalar-product}
for different sets, $\vec{x}\, \slashed{=}\,\vec{x}'$.

The eigenfunctions of the operator $B_N(u)$~\eqref{PsiB} diagonalize the Hamiltonian $\widetilde{\mathbb{H}}_N$~\eqref{HamB}.
The corresponding eigenvalues can be found either with the help of QISM machinery that involves  construction of the relevant
Baxter-$Q$ operators, see e.g. \cite{Derkachov:1999pz,Belitsky:2014rba}, or by a a more brute-force approach described below.
To this end we notice that in the region
$z_1 \gg z_2\gg\ldots \gg z_N$, i.e. $|z_{k+1}/z_k|= \mathcal{O}(\epsilon)$, $\epsilon\to 0$,  the eigenfunction~\eqref{PsiB} is simplified to
a linear combination of ``plane waves''
\begin{align}\label{asPsi}
\Psi_{\{p,\vec{x}\}}(z_1,\ldots,z_N)\sim c(p,\vec{x})\,{z_{1}^{-s+ix_{1}}\ldots z_{N-1}^{-s+ix_{{N-1}}}} e^{-ip z_N} + \cdots\,,
\end{align}
where $c(p,\vec{x})$ is a certain coefficient and
ellipses stand for the similar terms with permutations of the separated variables $x_1,\ldots, x_{N-1}$.
Action of the Hamiltonian~\eqref{HamB} on the eigenfunction~\eqref{asPsi} can be brought to a more convenient form using the following
identities~\cite{Bartels:2011nz,Belitsky:2014rba,Braun:2014owa}:
\begin{align}
\ln (i\partial_z) =\psi(-z\partial_z)-\ln (iz)\,, &&
\ln (iz^2\partial_z) =\psi(z\partial_z)+\ln (iz)\,.
\end{align}
The ``bulk'' Hamiltonians can be written as $\mathcal{H}_{kk+1}=h_{k,k+1}+h_{k+1,k}$
where~\cite{Derkachov:2005hw,Derkachov:2005fg}
\begin{align}
h_{kk'}=\psi\big( z_{kk'}\partial_k +2s\big)-\psi(1)\,.
\end{align}
These expressions can be simplified in the chosen kinematics as
\begin{align}
h_{k,k+1} &= \psi(z_k \partial_k+2s)-\psi(1)+ \mathcal{O}(\epsilon)\,,
\notag\\
h_{k+1,k} &= \ln (iz_k)+\psi(-z_{k+1} \partial_{k+1})-\psi(1)-\ln(iz_{k+1}) + \mathcal{O}(\epsilon)\,.
\end{align}
The last identity can be obtained as follows
\begin{align}
\psi\big( z_{k+1,k}\partial_{k+1} +2s\big) &=e^{-z_k\partial_{k+1}}z_{k+1}^{-2s}\psi(z_{k+1}\partial_{k+1}) z_{k+1}^{2s}
e^{z_k\partial_{k+1}}
\notag\\
&=
e^{-z_k\partial_{k+1}}(-\ln iz_{k+1} +z_{k+1}^{-2s}\ln i(z_{k+1}^2\partial_{k+1}) z_{k+1}^{2s})
e^{z_k\partial_{k+1}}
\notag\\
&= -\ln iz_{k+1,k} +\ln z_{k}^2+ \ln(i\partial_{k+1})+\mathcal{O}(\epsilon)
\notag\\
&=
\ln (iz_k)+\psi(-z_{k+1} \partial_{k+1})-\ln(iz_{k+1})+ \mathcal{O}(\epsilon)\,.
\end{align}
Thus, in the chosen region $|z_{k+1}/z_k|= \mathcal{O}(\epsilon)$,  the Hamiltonian $\widetilde {\mathbb{H}}_N$ takes the form
\begin{align}
\widetilde
{\mathbb{H}}_N=\sum_{k=1}^{N-1}\big (\psi(-z_k\partial_k)+\psi(z_k\partial_k+2s)-2\psi(1)\big)
 + 2\ln(i\mu\partial_N) - 2\psi(1)  + O(\epsilon)\,,
\end{align}
from which one immediately reads the eigenvalue from Eq.~\eqref{asPsi}
\begin{align}
{E}_{p,\vec{x}}=2\ln (\mu p) - 2 \psi(1) +\sum_{k=1}^{N-1}\Big(\psi(s+i x_k)+\psi(s-i{x_k})-2\psi(1)\Big)\,.
\end{align}
The special case considered in~\cite{Braun:2014npa} (heavy baryons) corresponds to $s=1$, $N=2$.
In this case one obtains (in a certain convenient normalization)
\begin{align}
 \Psi_{p,x}(z_1,z_2)&= p \int_0^1d\alpha\,\left(\frac{\alpha}{\bar\alpha}\right)^{ix}
\exp[{-}ip(\bar\alpha z_1+\alpha z_2)]\,,\qquad \bar\alpha = 1-\alpha\,,
\end{align}
and
\begin{align}
E_{p,x} = 2\ln(\mu p) + \psi(1+i x)+\psi(1-i x)+4\gamma_E -2\sigma_h-4\sigma_q \,,
\label{gamma}
\end{align}
where we have added the constant term corresponding to the (light and heavy) field anomalous dimensions.

Finally, one has to perform an inversion transformation $z\to -1/z$ \eqref{inversion} in order to go back to the original problem with
the Hamiltonian commuting with $C_2(u)$, and obtain the wave function, $\Psi_{p,x}(z_1,z_2)\stackrel{J}{\mapsto} \Phi_{p,x}(z_1,z_2)$
~\cite{Braun:2014npa}
\begin{align}\label{Psidef}
{\Phi}_{p,x}(z_1,z_2)&=\frac{p}{z_1^2 z_2^2}\int_0^1d\alpha\,\left(\frac{\alpha}{\bar\alpha}\right)^{ix}
\exp[ip(\bar\alpha/z_1+\alpha/z_2)]
\notag\\&
=
\frac{p \pi x}{\sinh(\pi x)} \frac{e^{ip/z_1}}{z^2_1z_2^2}\,{}_1F_1\Big(1\!+\!ix,2,ip \big(z^{-1}_2\!-\!z^{-1}_1\big)\Big),
\end{align}
which coincides with the expression found in~\cite{Braun:2014npa}.
Note that in this representation $p$ ($p\mapsto s$ in the notation of~\cite{Braun:2014npa}) becomes an eigenvalue of the generator of
special conformal transformations
\begin{align}
  \big( S_+^{(1)}+ S_+^{(2)} \Big) \Phi_{p,x}(z_1,z_2) = -i p\, \Phi_{p,x}(z_1,z_2)\,.
\end{align}
The functions \eqref{Psidef} define the basis of states for the heavy-light baryon DAs with
autonomous scale dependence (for the case of aligned light-quark helicities), where $z_1$ and $z_2$
are the light-quark light-cone coordinates.
The energies \eqref{gamma} are nothing but the corresponding anomalous dimensions.

\subsection{Open spin chains}
A systematic approach to construct integrable models with nontrivial boundary conditions (open spin chains) was developed by
Sklyanin~\cite{Sklyanin:1988yz}. The monodromy matrix for such systems is given by the following
expression~\footnote{Our definition differs from the standard one,
$\mathbb{T}_N(u)  = T_N(u) T_N^{-1}(-u+i)$, by a numerical factor.}
\begin{align}\label{monodromy-open}
\mathbb{T}_N(u)
= (-1)^N\,T_N(u) \sigma_2 T^{t}_N(-u) \sigma_2 =
\begin{pmatrix}
\mathbb{A}_N(u) & \mathbb{B}_N(u)\\
\mathbb{C}_N(u) & \mathbb{D}_N(u)
\end{pmatrix},
\end{align}
where $T_N(u)$ is the monodromy matrix of the closed spin chain~\eqref{monodromy-closed}, $T^t_N(u)$ is the transposed matrix.
It can be written in terms of the Lax operators as follows:
\begin{align}\label{T-LL}
\mathbb{T}_N(u) =L_1(u) L_2(u)\ldots L_N(u)\, L_N(u)\ldots  L_2(u) L_1(u)\,.
\end{align}
This representation can easily be obtained  using the identity
\begin{align}\label{T-Lax}
(L_k(u))^t= -\sigma_2 L_k(-u)\sigma_2.
\end{align}
Off-diagonal
elements of the monodromy matrix satisfy the following relations (see e.g. \cite{Derkachov:2003qb})
\begin{align}
{(-2u+i)}\,{\mathbb{B}_N(-u)}=(2u+i)\,{\mathbb{B}_N(u)}\,, &&{(-2u+i)}\,{\mathbb{C}_N(-u)}={(2u+i)}\,{\mathbb{C}_N(u)}\,.
\end{align}
As a consequence, the operators $\mathbb{B}_N(u)$ and $\mathbb{C}_N(u)$  vanish at $u=i/2$ and can be represented as
$\mathbb{B}_N(u)=(-2u+i) \widehat{ \mathbb{B}}_N(u)$ and similarly for $\mathbb{C}_N(u)$.
The operators with a ``hat''  are even functions of $u$,
 $\widehat{ \mathbb{B}}_N(u)=\widehat{ \mathbb{B}}_N(-u)$,  $\widehat{ \mathbb{C}}_N(u)=\widehat{ \mathbb{C}}_N(-u)$.
These operators and the transfer matrix,
\begin{align}
\mathbf{t}_N(u)= {\mathbb{A}}_N(u) +{ \mathbb{D}}_N(u),
\qquad \mathbf{t}_N(u)=\mathbf{t}_N(-u),
\end{align}
 form  commuting operator families~\footnote{Note that in distinction with the closed spin chains diagonal elements of
the monodromy matrix for open spin chains do not commute.}
\begin{align}
 [\widehat{ \mathbb{B}}_N(u),\widehat{ \mathbb{B}}_N(v)]\,=\,
[\widehat{ \mathbb{C}}_N(u),\widehat{ \mathbb{C}}_N(v)]\,=\,
 [\mathbf{t}_N(u),\mathbf{t}_N(v)]\,=\,0\,.
\end{align}

Our aim is to construct a Hamiltonian which commutes with $\widehat{\mathbb{C}}_N(u)$ (or with $\widehat{\mathbb{B}}_N(u)$).
Let us consider at first the homogeneous spin chain. In this case using the representation in~\eqref{T-LL} for the
monodromy matrix and the relation in~\eqref{HLL},
it is easy to find that the commutator of the \textit{bulk} Hamiltonian
\begin{align}\label{open-H-bulk}
\mathbf{H}_{N}=\sum_{k=1}^{N-1}\mathcal{H}_{kk+1},\qquad \mathcal{H}_{kk+1}=2(\psi(J_{kk+1})-\psi(1))\,,
\end{align}
and the monodromy matrix reads
\begin{align}\label{open-HT}
[\mathbf{H}_{N},{\mathbb T}_{N}(u)]
&=-i L_2(u)\ldots L_N(u)\, L_N(u)\ldots  L_1(u) +
                                   i L_1(u) \ldots L_N(u)\, L_N(u)\ldots  L_2(u)
\notag\\ &~
    = i[L_1(u), {\mathbb T}_{N-1}(u)]\,,
\end{align}
where ${\mathbb T}_{N-1}(u)= L_2(u)\ldots L_{N}(u) L_{N}(u)\ldots L_2(u)$. Taking the trace over the auxiliary space in
\eqref{open-HT} results in
\begin{align}
[\mathbf{H}_N, \mathbf{t}_N(u)]=0\,.
\end{align}
Furthermore, considering the off-diagonal matrix elements of \eqref{open-HT} and  taking into account~Eq.~\eqref{SS+} one can
show that
\begin{align}\label{BCopenH}
[\ln\big(i\mu S_+^{(1)}\big)+\mathbf{H}_N, \widehat{\mathbb{C}}_N(u)]=0 &&\text{and}
&& [\ln\big(-i\mu S_-^{(1)}\big)+\mathbf{H}_N, \widehat{\mathbb{B}}_N(u)]=0\,.
\end{align}
Thus in the homogeneous case the Hamiltonian belonging to the $\mathbb{C}_N(u)$ family is obtained from the bulk
Hamiltonian~\eqref{open-H-bulk}  by adding the boundary operator $\ln\big(i\mu S_+^{(1)}\big)$.
In full analogy to the closed spin chain it can be shown that the operators
$ \widehat{\mathbb{C}}_N(u) $ and $ \widehat{\mathbb{B}}_N(u)$ are related to each other by the inversion transformation,
$ J\widehat{\mathbb{C}}_N(u)=-\widehat{\mathbb{B}}_N(u)J$.
Therefore, it is sufficient to consider only one of them.

\subsubsection{Inhomogeneous chains}\label{sect:inhomogeneous}

Spin systems that are interesting in QCD context are somewhat more complicated and correspond to
inhomogeneous open spin chains with impurities~\cite{Derkachov:1999ze}. In a typical setup
one can assume that the  spins on all sites except for the last one are equal to each other,
$s_k=s$, $1\leq k<N$.
The monodromy matrix in this case takes the form
\begin{equation}\label{TL-explicit}
{\mathbb T}_N^{(w)}(u)=L_1(u) L_2(u)\ldots L_N(u+iw)\, L_N(u-iw)\ldots  L_2(u) L_1(u),
\end{equation}
where $w$ is an impurity parameter. The dependence on $w$ comes only through the Lax operators $L_N$ so that it is easy to
check that ${\mathbb T}^{(w)}_N(u)$ is an even function of $w$, i.e. $\widehat{\mathbb{B}}^{(w)}_N(u) =
\widehat{\mathbb{B}}^{(-w)}_N(u)$ etc. Since $(\widehat{\mathbb{B}}^{(w)}_N(u))^\dagger = \widehat{\mathbb{B}}^{(-w^*)}_N(u^*)
= \widehat{\mathbb{B}}^{(w^*)}_N(u^*)$,   
this symmetry implies that the corresponding conserved charges are hermitian operators
 if the impurity parameter $w$ is either \textit{real} or  \textit{imaginary}.

We will show that the operators $\widehat{\mathbb{B}}^{(w)}_N(u)$,  $\widehat{\mathbb{C}}^{(w)}_N(u)$ commute with
the Hamiltonian ${{\mathscr{H}}_N^{(w)}} $ with added boundary operators, cf.~\eqref{BCopenH},
\begin{align}\label{BCopenH1}
[\ln\big(i\mu S_+^{(1)}\big)+{{\mathscr{H}}_N^{(w)}}, \widehat{\mathbb{C}}_N(u)]=0\,,
&&
[\ln\big(-i\mu S_-^{(1)}\big)+{{\mathscr{H}}_N^{(w)}}, \widehat{\mathbb{B}}_N(u)]=0\,,
\end{align}
which is modified compared to $\mathbf{H}_N$ \eqref{open-H-bulk} by one term
\begin{align}\label{openH}
  {{\mathscr{H}}_N^{(w)}}=\sum_{k=1}^{N-2} \mathcal{H}_{kk+1} + {\mathcal{H}}^{(w)}_{N-1N}\,,
\end{align}
where
\begin{align}
\mathcal{H}_{kk+1} & =  R_{kk+1}^{-1}(0) \frac{d}{du} R_{kk+1}(iu)\Big|_{u=0} = 2\Big(\psi(J_{kk+1})-\psi(1)\Big)\,,
\\
{\mathcal{H}}^{(w)}_{N-1N} & =
R_{N-1N}^{-1}(iw) \frac{d}{dw} R_{N-1N}(iw)
\notag\\&=
\psi(J_{N-1N}+w)+\psi(J_{N-1N}-w) -\psi(1+w)-\psi(1-w)\,.
\end{align}
Here $R_{kk+1}$ is the $sl(2)$-invariant $R$-matrix~\cite{Kulish:1981gi}
\begin{align}
R_{kk+1}(u)=(-1)^{J_{kk+1}-s_k-s_{k+1}}\frac{\Gamma(J_{kk+1}-iu)}{\Gamma(J_{kk+1}+iu)}\frac{\Gamma(1+iu)}{\Gamma(1-iu)}\,.
\end{align}
Note that $R_{kk+1}^{-1}(u)=R_{kk+1}(-u)$, and hence ${\mathcal{H}}^{(w)}_{N-1N}$ is a hermitian operator for both
real and imaginary $w$.

The main task is to calculate the commutator of the Hamiltonian ${{\mathscr{H}}_N^{(w)}} $ with the monodromy matrix ${\mathbb T}^{(w)}_N(u)$.
Using the relation in~\eqref{HLL} repeatedly one obtains
\begin{align}\label{open-HT-im}
\left[\sum_{k=1}^{N-2}
\mathcal{H}_{kk+1}, { \mathbb T}^{(w)}_N(u)\right] & = i[L_1(u), {\mathbb T}^{(w)}_{2\to N}(u)]
+i L_{1\cdots N-2}\Big\{ L_{N}(u+iw) L_{N}(u-iw) L_{N-1}(u)
\notag\\
&\quad -
 L_{N-1}(u)L_{N}(u+iw) L_{N}(u-iw)\Big\}L_{N-2\cdots 1},
\end{align}
where $L_{1\cdots M}=L_1(u)\cdots L_{M}(u)$ and ${\mathbb T}^{(w)}_{2\to N}(u)= L_2(u)\ldots L_{N}(u+iw) L_{N}(u-iw)\ldots L_2(u)$.
The expression in the curly brackets on the  r.h.s. of Eq.~\eqref{open-HT-im} can be simplified to
\begin{align}\label{X0}
\{\ldots\}=(2u-i) \Big\{L_N(u) L_{N-1}(u)-L_{N-1}(u)L_N(u) \Big\}
\end{align}
with the help of the following identity for the Lax operators:
\begin{align}\label{L2}
L_k^2(u)=\rho_{s}(u)+\left(2u-i\right)L_k(u)\,, && \rho_{s}(u)=-(s+iu)(s-1-iu)\,.
\end{align}
{Next, differentiating the $RLL$ relations (below $R(iw)\equiv R_{N-1,N}(iw)$):}
\begin{align}\label{RLLw}
 R(iw) L_{N-1}(u) L_{N}(u-iw) & = L_{N}(u-iw)L_{N-1}(u)R(iw)\,,
 \notag\\
 R(iw) L_{N}(u+iw) L_{N-1}(u) & = L_{N-1}(u)L_{N}(u+iw)R(iw)
\end{align}
with respect to $w$ results in
\begin{align}
[{\mathcal{H}}^{(w)}_{N-1N}, L_{N}(u-iw)L_{N-1}(u)]&=i\left(R(iw)L_{N-1}(u)R(-iw)-L_{N-1}(u)\right),
\notag\\
[{\mathcal{H}}^{(w)}_{N-1N},L_{N-1}(u)L_{N}(u+iw)]&=i\left(L_{N-1}(u)-R(iw)L_{N-1}(u)R(-iw)\right).
\end{align}
Thus we get
\begin{align}\label{HLX}
 [{\mathcal{H}}^{(w)}_{N-1N}, { \mathbb T}^{(w)}_N(u)] = i L_{1\cdots N-2}\, \mathbb{X} \, L_{N-2\cdots 1}\,,
\mathcal{}\end{align}
where
\begin{align}
\mathbb{X }& = -i[{\mathcal{H}}^{(w)}_{N-1N}, L_{N-1}(u)L_{N}(u+iw)L_{N}(u-iw)L_{N-1}(u)]
\notag\\
& = -2iw L_{N-1}^2(u) -R(iw)\Big( L_{N-1}^2(u)L_N(u-iw) - L_N(u+iw) L_{N-1}^2(u)\Big) R(-iw).
\end{align}
Using Eqs.~\eqref{L2} and \eqref{RLLw} this expression can be simplified to
\begin{align}
\mathbb{X}=-(2u-i) \Big(L_N(u) L_{N-1}(u)-L_{N-1}(u)L_N(u) \Big)\,,
\end{align}
which exactly cancels \eqref{X0}. Thus we obtain
\begin{align}\label{THN}
[{{\mathscr{H}}_N^{(w)}},  { \mathbb T}^{(w)}_N(u)]= i[L_1(u),{ \mathbb T}^{(w)}_{2\to N}(u)].
\end{align}
Finally, adding the contribution from the boundary operator $\ln (-i\mu S_-^{(1)})$ one
ends up with the desired result
\begin{align}
[\ln\big(-i\mu S_-^{(1)}\big) + {{\mathscr{H}}_N^{(w)}}, \widehat{\mathbb{B}}^{(w)}_N(u)]=0\,.
\end{align}
The similar equation with $\ln\big(-i\mu S_-^{(1)}\big)\mapsto \ln\big(i\mu S_+^{(1)}\big)$ holds for $\widehat{\mathbb{C}}^{(w)}_N(u)$.

It is instructive to compare this result with the more conventional spin chains that appear in the
analysis of the RGE for light quark-gluon operators, of the type $\chi(z_1n)  f(z_2n)\ldots f(z_{N-1}n) \psi(z_Nn)$, cf.~\eqref{examples}.
In such applications the spins on the first and the last sites can differ from the spins in the bulk, which are all equal,
and also the impurity parameters $\omega_1$ and $\omega_N$, on the first and the last site can be nonzero.
The monodromy matrix for such a system takes the form
\begin{align}\label{TMLight}
 { \mathbb T}^{(w_1,w_N)}_N(u)=L_1(u+iw_1)L_2(u)\ldots L_N(u+iw_N) L_N(u-iw_N)\ldots L_2(u) L_1(u-iw_1)\,.
\end{align}
The conserved charges are generated by the transfer matrix,
$\mathbf{t}_N^{(w_1,w_N)}(u)= \text{Tr}\, {\mathbb T}^{(w_1,w_N)}_N(u)$, which commutes with the
Hamiltonian~\cite{Derkachov:1999ze}
\begin{align}
{\mathscr{H}}^{(w_1,w_N)}_N= {\mathcal{H}}_{12}^{(w_1)}+\sum_{k=2}^{N-2}
\mathcal{H}_{kk+1} + {\mathcal{H}}_{N-1N}^{(w_N)}\,.
\end{align}
The proof given in~\cite{Derkachov:1999ze} is not explicit. A more direct way to show that
$[{\mathscr{H}}^{(w_1,w_N)}_N,\mathbf{t}_N^{(w_1,w_N)}(u)]=0$ is the following.
Making use of Eq.~\eqref{THN} we get
\begin{align}\label{1/2Comm}
\Big[\sum_{k=2}^{N-2}
\mathcal{H}_{kk+1} + {\mathcal{H}}^{(w_N)}_{N-1N},\mathbf{t}_N^{(w_1,w_N)}(u)\Big] & =i\text{Tr}\left( L_1(u+iw_1)
\Big[L_2(u),
 { \mathbb T}^{(w_N)}_{3\to N}(u)\Big] L_1(u-iw_1)\right)
 \notag\\
&=i\text{Tr}\left( L_1(u)\Big[L_2(u),
 { \mathbb T}^{(w_N)}_{3\to N}(u)\Big] L_1(u)\right)\,.
\end{align}
In order to calculate the commutator with the remaining term, ${\mathcal{H}}_{12}^{(w_1)}$, it is convenient to
write the transfer matrix in a different form.
Namely, using Eq.~\eqref{T-Lax} 
one can write
$\mathbf{t}_N^{(w_1,w_N)}(u)= \text{Tr}\, \widetilde{\mathbb T}^{(w_1,w_N)}_N(-u)$, where
\begin{align}\label{TMLightN1}
 {\widetilde{ \mathbb T}^{(w_1,w_N)}_N}(u)=L_N(u-iw_N)\ldots L_2(u) L_1(u-iw_1) \,L_1(u+iw_1)L_2(u)\ldots  L_N(u+iw_N)\,.
\end{align}
Using this representation it is easy to bring the commutator in question to a form similar to~\eqref{HLX}
\begin{align}\label{HLX1}
[{
\mathcal{H}}_{12}^{(w_1)}, {\widetilde{\mathbb T}^{(w_1, w_N)}_N}(u)] = i L_{N\cdots 3}\, \mathbb{X}' \, L_{3\cdots N}\,,
\mathcal{}\end{align}
with
\begin{align}
\mathbb{X}'=-(2u-i) \Big(L_1(u) L_{2}(u)-L_{2}(u)L_1(u) \Big)=-\Big(L_1^2(u) L_{2}(u)-L_{2}(u)L_1^2(u) \Big)\,.
\end{align}
Finally, using~\eqref{T-Lax} one can bring the trace of \eqref{HLX1} to the same form as in Eq.~\eqref{1/2Comm} with an
opposite sign, which completes the proof.

\section{Heavy-light eigenfunctions}\label{sec:Eigenfunctions}

The basis formed by the eigenfunctions of the operator $\widehat{\mathbb{B}}_N$
plays a distinguished role in the QISM and
defines the so-called Sklyanin representation of separated variables (SoV)~\cite{Sklyanin:1995bm}.
Since $\widehat{\mathbb{B}}_N$ commutes with the Hamiltonian \eqref{BCopenH1}, the latter is diagonalized in this
basis and therefore the calculation of its spectrum becomes straightforward.
These eigenfunctions  for homogeneous spin chains, for closed as well as open ones, were constructed
in~\cite{Derkachov:2002tf,Derkachov:2003qb}. For our applications it is necessary to generalize this
construction to inhomogeneous open spin chains with impurities.

\begin{figure}[t]
\centerline{\includegraphics[width=0.450\textwidth]{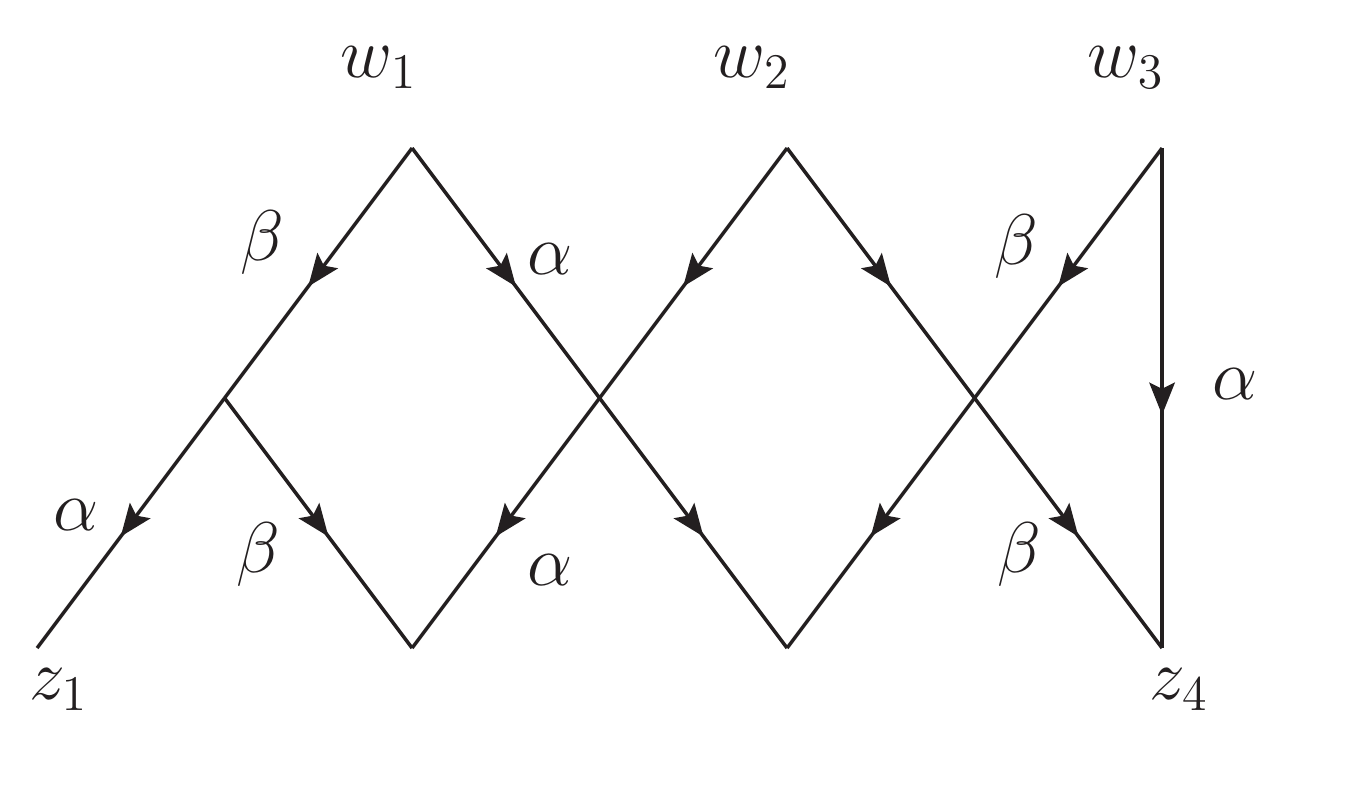}}
\caption{Diagrammatic representation of the layer operator $\Lambda_{N=4}(x)$ for open spin chains.
The notations are the same as in Fig.~\ref{fig:Lambda_closed}, the vertices imply integration with the measure~\eqref{measure} .
}
\label{fig:Lambda_open}
\end{figure}

The eigenfunctions of ${B}_N$  operator
for homogeneous closed spin chains were already introduced in Sec.~\ref{sec:closed}, Eq.~\eqref{PsiB}.
The  eigenfunctions for the open spin chain~\cite{Derkachov:2002tf,Derkachov:2003qb} take a  similar form.
In both cases the eigenfunctions are labeled by $N$ real parameters, $X=\{p,x_1,\ldots,x_{N-1}\}$,
where $p\geq 0$,  and have the  form
\begin{align}\label{PsiBoth}
\Psi_X(\vec{z})\equiv
\Psi_{\{p,x_1,\ldots,x_{N-1}\}}(\vec{z})= b_N(p)
\Lambda_N(x_1)\Lambda_{N-1}(x_2)\ldots \Lambda_{2}(x_{N-1}) e^{-ip z}\,.
\end{align}
Here $\vec{z}=\{z_1,\ldots,z_N\}$ and $b_N(p)$ is a normalization coefficient which we choose as
\begin{align}
 b_N(p)&=p^{Ns-1/2} (\Gamma(2s))^{-N^2/2} &&\text{closed~chain}\,,
\notag\\
 b_N(p)&=p^{Ns-1/2} (\Gamma(2s))^{-N(N-1/2)} &&\text{open~chain}\,.
\end{align}
All differences between the closed and open spin chain in the construction~\eqref{PsiBoth} come from the form of the layer operators
$\Lambda_M(x)$,
\begin{align}
[\Lambda_M f](z_1,\ldots,z_M)=\left(\prod_{k=1}^{M-1}\int D_s w_k\right)
\Lambda_M(z_1,\ldots,z_M| w_1,\ldots w_{M-1}) f(w_1,\ldots,w_{M-1}).
\end{align}
The function $\Lambda_M(z_1,\ldots,z_M| w_1,\ldots w_{M-1})$ for a closed chain is given in Eq.~\eqref{closedL} and can be
visualized as the diagram in Fig.~\ref{fig:Lambda_closed}.
The corresponding expression for an open chain is more complicated and is presented in diagrammatic form (for $M=4$)
in Fig.~\ref{fig:Lambda_open} where the vertices imply integration with the measure~\eqref{measure}.
%
%
%
\begin{figure}[t]
\centerline{\includegraphics[width=0.87\textwidth]{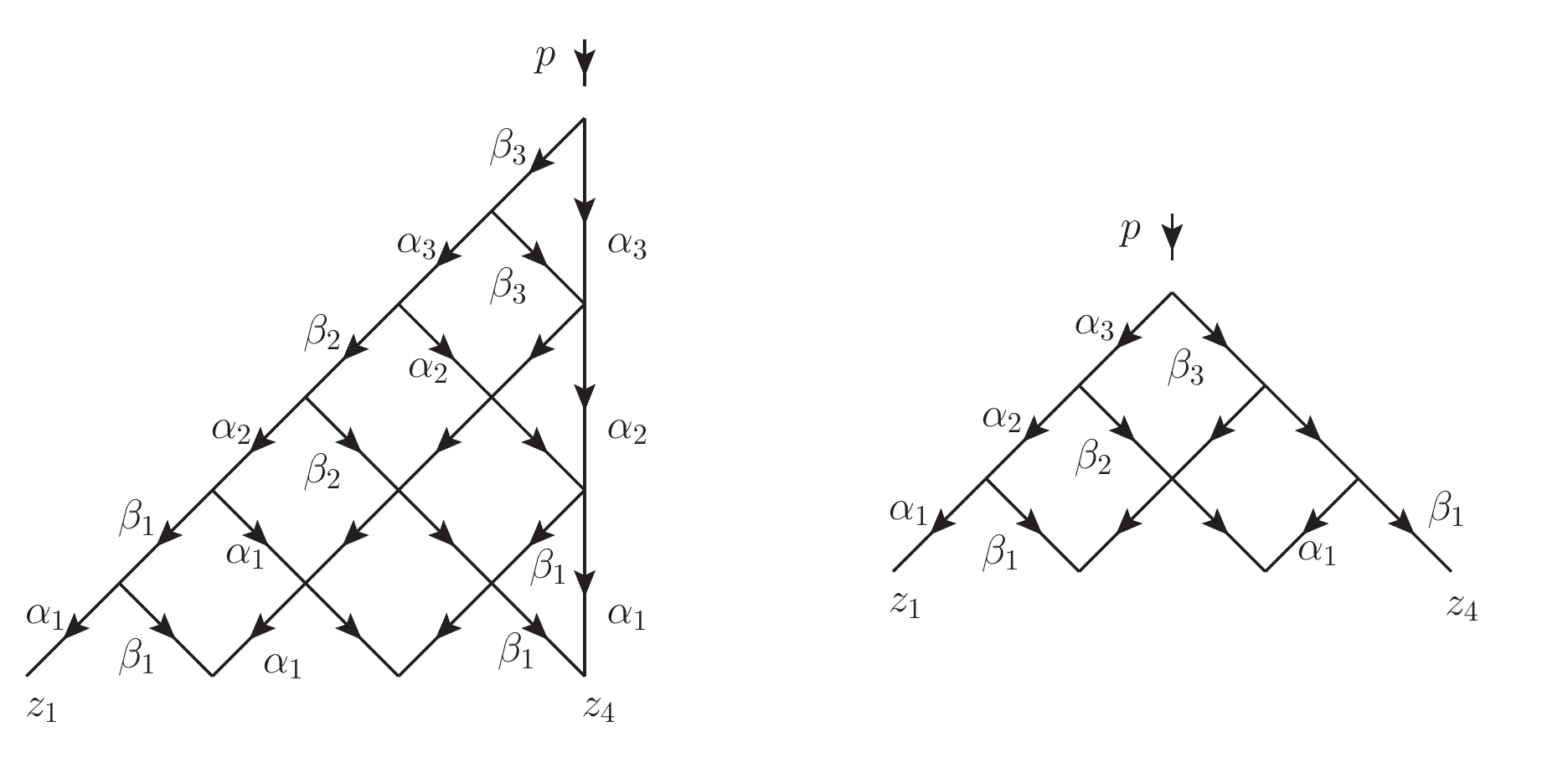}}
\caption{Diagrammatic representation for the eigenfunctions of the open (left) and closed (right) homogeneous
spin chains for $N=4$.
The indices are $\alpha_k=s-ix_k$ and $\beta_k=s+ix_k$. }
\label{fig:open-closed}
\end{figure}
%
Diagrammatic representation for the eigenfunctions  $\Psi_X(\vec{z})$ of open and closed spin chains
is shown in Fig.~\ref{fig:open-closed}.

The proof that $\Psi_X(\vec{z})$~\eqref{PsiBoth} diagonalizes the corresponding $B$-operator relies on the following
properties of the layer operators~\cite{Derkachov:2002tf,Derkachov:2003qb}:
\begin{align}\label{Lambda-properties}
B_N(x)\Lambda_N(x)=0 && \text{and}  &&\Lambda_M(x)\Lambda_{M-1}(y)=\Lambda_M(y)\Lambda_{M-1}(x)\,.
\end{align}
 The layer operator for open chain has an additional symmetry $\Lambda_k(x)=\Lambda_k(-x)$.
 From the second equation in~\eqref{Lambda-properties} it is obvious that $\Psi_{X}$ is a symmetric function of
 separated variables $x_1,\ldots,x_{N-1}$ ($x_1^2,\ldots,x_{N-1}^2$ for the open chain).
 The first equation  ensures that the operator $B_N(x_k)$ (for open chain $B_N(\pm x_k)$) annihilates
 the function $\Psi_X$. Taking into account that the operator $B_N(u)$ (closed chain)
 is a polynomial of degree $N-1$ in $u$ and the coefficient of the senior power $u^{N-1}$
 is simply $S_-^{(1)}+\ldots+ S_-^{(N)}$, one can write it in the form 
 \begin{align}
B_N(u)= i{\left(\sum_{k=1}^NS_-^{(k)}\right)} \prod_{k=1}^{N-1}(u-x_k)+\sum_{j=1}^{N-1}\left(\prod_{k\neq j}\frac{u-x_k}{x_j-x_k}\right) B_N(x_j)\,.
\label{BNrep}
\end{align}
From this representation using Eq.~\eqref{pEigen} it follows immediately
that $\Psi_{X}$ for closed chain
is an eigenfunction of the operator $B_N(u)$ with an eigenvalue $b_N(u)=-p\prod_{k=1}^{N-1}(u-x_k)$.
Similarly, the operator $\widehat {\mathbb{B}}_N(u)$ for open chain is a polynomial of degree $N-1$ in $u^2$ and
the representation analogous to \eqref{BNrep} holds, with obvious substitutions $u \to u^2$, $x_j\to x_j^2$.
It follows that $\Psi_{X}$ for open chain is an eigenfunction of $\widehat {\mathbb{B}}_N(u)$
with an eigenvalue $b_N(u)=-p\prod_{k=1}^{N-1}(u^2-x^2_k)$.

Since $B_N(u)$ (closed chain) and $\widehat {\mathbb{B}}_N(u)$ (open chain) are self-adjoint operators for real $u$,
their eigenfunctions are orthogonal for different sets $X, X'$ of  separated variables. One
obtains~\cite{Derkachov:2002tf,Derkachov:2003qb}
\begin{align}\label{s-product-B}
{{\VEV{\Psi_{X'}, \Psi_{X}}}
}
=(2\pi)^{N-1}\,
\,\delta(p-p')\,
\left(\sum_{S} \delta^{(N-1)}(\vec{x}-S\vec{x}')\right)\,
\frac{\prod_{j\neq k}
\Gamma(i(x_k-x_j))}{\prod_{k=1}^{N-1}\big[\Gamma(\alpha_{x_k})\Gamma(\beta_{x_k})\big]^N}\,,
\end{align}
and
\begin{align}\label{s-product-open}
{{\VEV{\Psi_{X'},\Psi_{X}}}
} &=(2\pi)^{N-1}\,
\,\delta(p-p')\,
\left(\sum_{S} \delta^{(N-1)}(\vec{x}-S\vec{x}')\right)\,
\notag\\
&\quad \times
\prod_{n=1}^{N-1}\Gamma(2ix_n) \Gamma(-2ix_n)\,
\frac{\prod_{j< k}
\Gamma(i(x_k\pm x_j))\Gamma(-i(x_k\pm x_j)) }{\prod_{k=1}^{N-1}\left[\Gamma(\alpha_{x_k})\Gamma(\beta_{x_k})\right]^{2N}}\,,
\end{align}
for closed and open chain, respectively.
In these expressions $\alpha_{x}=s-ix$, $\beta_{x}=s+ix$, the sum $\sum_{S}$ goes over all possible permutations of
separated variables $\vec{x}' = \{x_1',\ldots, x_{N-1}'\}$,
and for the open spin chain it is assumed that all $x_k\in \mathbb{R}_+$.

In the applications to operator renormalization in HQET~\cite{Braun:2015pha,Braun:2017liq} one encounters
an inhomogeneous open spin chain with impurities and the above construction of the eigenfunctions
of  $\widehat{\mathbb{B}}_N$ operator {for homogeneous spin chain} has to be modified. In practice one needs a special case where all spins
except for the last one are equal, $s_k=s$ for $k<N$ while $s_N\neq s$, and the impurity parameter $\xi_N\equiv i\omega \neq 0$.

\begin{figure}[t]
\centerline{\includegraphics[width=0.47\textwidth]{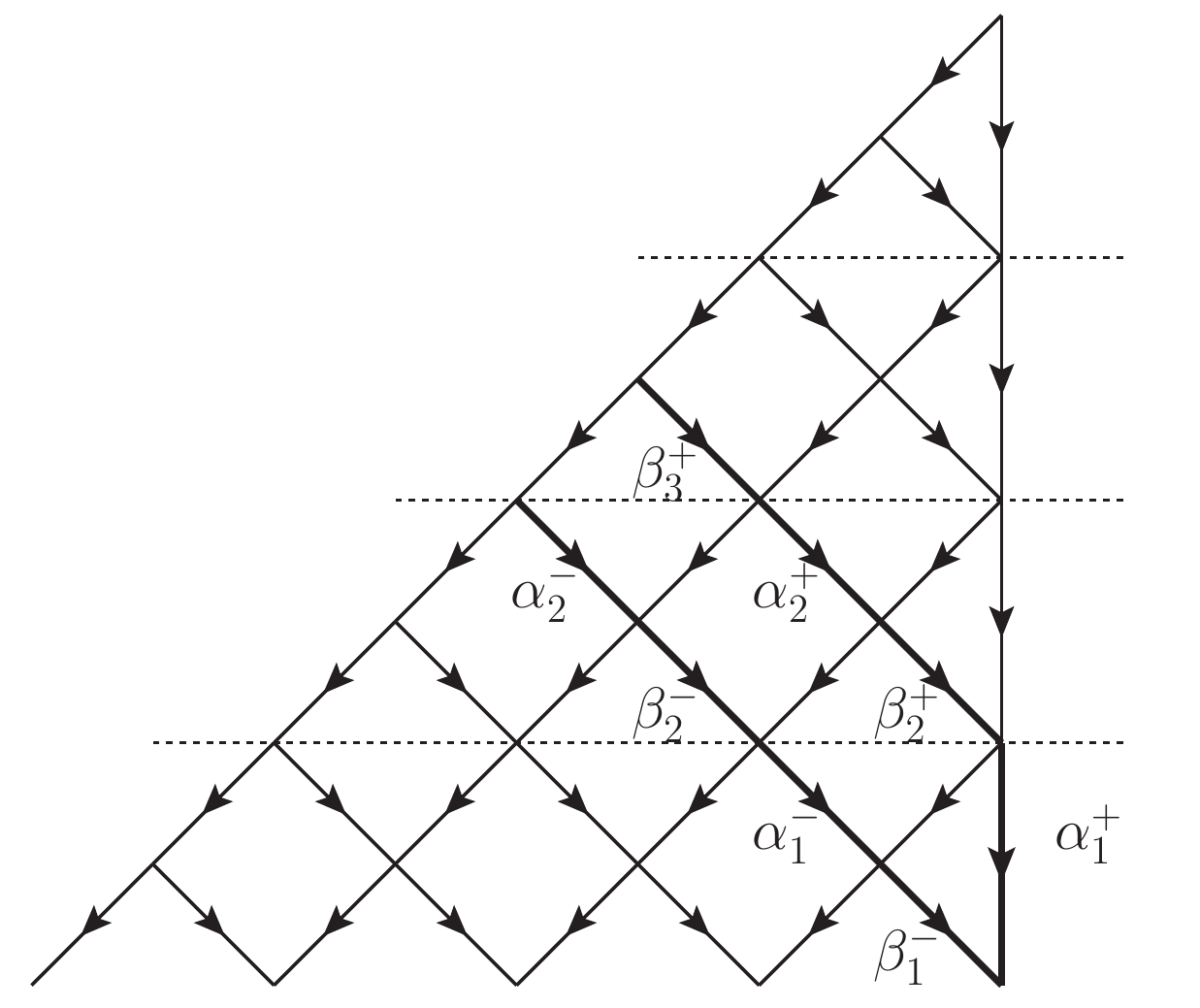}}
\caption{Diagrammatic representation for an eigenfunction of the operator $\widehat{\mathbb{B}}_N$ for an inhomogeneous
open spin chain for $N=5$. The modified propagators are shown by fat lines with arrows.
The corresponding modified indices  are $\alpha^\pm_k=s_N\pm w-ix_k$ and $\beta_k^\pm=s_N\pm w+ix_k$. The conformal spin in the integration measure
in the vertices is always given by the half sum of the indices of the {\it outgoing}  lines. The dashed lines separate different layers.}
\label{fig:B-w}
\end{figure}

Note that the operator ${\widehat{\mathbb{B}}^{(w)}_N(u)}$ depends only on $w^2$, as can easily be seen from Eq.~\eqref{TL-explicit},
and as a consequence ${\widehat{\mathbb{B}}_N^{(w)}(u)}$ is a self-adjoint operator for both real and imaginary $w$.
One can show that the corresponding eigenfunction has the
form~\eqref{PsiBoth} with modified layer operators that still obey the relations in~\eqref{Lambda-properties}. The necessary
changes are summarized in Fig.~\ref{fig:B-w} where the modified propagators are shown by fat lines. Note that also the
integration measure for the vertices involving these lines is affected. Using the technique developed
in~\cite{Derkachov:2002tf,Derkachov:2003qb} it is easy to check  that the modified layer operators have
the required properties.

As mentioned above, the operator ${\widehat{\mathbb{B}}_N^{(w)}}$ depends on $w^2$ only so that
one expects $\Psi_{X}^{(w)}(\vec{z}) \simeq \Psi_{X}^{(-w)}(\vec{z})$. This symmetry is, however, not manifest as
the diagram for the eigenfunction in Fig.~\ref{fig:B-w} does not go into itself for $w\to -w$.
For $N=2$ going over to the momentum representation one can find explicit expression for the eigenfunction that is an even function
of $w$. For a general case, taking into account that the dependence on $w$ comes only
from the {$L_N(u \pm iw)$} operator, cf. Eq.~\eqref{TL-explicit}, one can argue that  $\Psi_{X}^{(w)}(\vec{z})$ and  $\Psi_{X}^{(-w)}(\vec{z})$
have to be proportional to one another
 and, comparing the asymptotics
in the region $z_1\gg z_2\gg\ldots\gg z_{N}$, find that indeed $\Psi_{X}^{(w)}(\vec{z})= \Psi_{X}^{(-w)}(\vec{z})$.
A manifestly symmetric representation for $N>2$ is  not known.

Assuming that $w$ is either small and positive, $0 < w < s_N$, or imaginary, one gets for the scalar product
\begin{align}\label{s-product-open-w}
{\VEV{\Psi^{(w)}_{X'},\Psi_{X}^{(w)}}
} &=(2\pi)^{N-1}\,\,\delta(p-p')\,
\left(\sum_{S}\delta^{(N-1)}(\vec{x}-S\vec{x}')\right)\,\prod_{n=1}^{N-1}{\Gamma(2ix_n) \Gamma(-2ix_n)}
\notag\\
&\quad \times
\frac{\prod_{j< k}
\Gamma(i(x_k\pm x_j))\Gamma(-i(x_k\pm x_j)) }{\prod_{k=1}^{N-1} {|\Gamma(s_N+w+ix_k)\Gamma(s_N-w+ix_k)|}^2
\left[\Gamma(\alpha_{x_k})\Gamma(\beta_{x_k})\right]^{2(N-1)}}\,.
\end{align}
In this case the set of functions $\Psi_{X}^{(w)}$ for real separated coordinates  $x_k$ form a complete system
and the operator ${\widehat{\mathbb{B}}_N^{(w)}}$ has only continuous spectrum.

If $w$ is real and $w\to s_N$, the $\Gamma$-functions in the denominator, $\Gamma(s_N-w+ix_k)$, develop singularities that
signal the formation of discrete states for $w>s_N$ corresponding to imaginary values of separated variables
~\footnote{Note that $\Psi^{(w)}_X$ is symmetric to $x_k\leftrightarrow-x_k$ and therefore it suffices to consider $\Im(x_k)>0$.}
\begin{align}
s_N-\omega-ix_k^{(n)}=-n\,, \text{\ where \ }  n \in \mathbb{N}\,, \qquad 0\leq n< \omega-s_N\,.
\end{align}
The complete description of the discrete eigenstates goes beyond the scope of this paper.
As an illustration we consider the $N=2$ case~\cite{Braun:2015pha,Braun:2017liq}, $s\mapsto s_1$, $s_{N=2} \mapsto s_2$.
To this end it is convenient to go over to the momentum space.
We define the momentum-space eigenfunction $\widetilde \Psi_{X}^{(w)}(p_1,p_2)$ by
\begin{align}
\Psi_{X}^{(w)}(z_1,z_2) &=  \int_0^\infty dp_1 \,p_1^{2s_1-1} \int_0^\infty dp_2\, p_2^{2s_2-1}
\, e^{-ip_1z_1-ip_2 z_2}\, \widetilde \Psi_{X}^{(w)} (p_1,p_2)\,,
\notag\\[3mm]
\widetilde \Psi_{X}^{(w)}(p_1,p_2) &
=\frac1{\Gamma(2s_1)\Gamma(2s_2)} {\VEV{e^{-ip_1z_1-ip_2 z_2} | \Psi_{X}^{(w)}(z_1,z_2)}_{s_1,s_2}}\, ,
\end{align}
where we have used
\begin{align}
\vev{e^{-ipz}\big|e^{-ip'z}}_{s}=\Gamma(2s)\,p^{1-2s}\,\delta(p-p')\, .
\end{align}
Using \eqref{PsiBoth} and evaluating the scalar product one obtains after a short calculation
(here we put the normalization constant $b_2(p)\to 1$)
\begin{align}
\label{example11}
\widetilde \Psi_{X}^{(w)}(p_1,p_2)&=p\,\delta(p-p_1-p_2)
\frac{\Gamma(s_1+s_2+\omega)\Gamma(s_1+s_2-\omega)}{\Gamma(2s_2)\Gamma(s_2+ix)\Gamma(s_2-ix)}\,
\notag\\
&\quad \times \left(\frac{p_1}{p_1+p_2}\right)^{w-s_1-s_2}
{}_2F_1\left(\genfrac{}{}{0pt}{}{s_2-\omega-ix,s_2-\omega+ix}{2s_2}\Big|-\frac{p_2}{p_1}\right)\,.
\end{align}
The scalar product \eqref{s-product-open-w} in the momentum space takes the form
\begin{align}
 \vev{ \Psi_{X}^{(w)},\widetilde \Psi_{X'}^{(w)}} &=
\Gamma(2s_1)\Gamma(2s_2)\int_0^\infty dp_1 dp_2 \, p_1^{2s_1-1}p_2^{2s_2-1}
\widetilde \Psi_{X}^{(w)}(p_1,p_2)\, \big({\widetilde \Psi}_{X'}^{(w)}\big)^\ast(p_1,p_2)
\,,
\end{align}
and using the explicit expression in \eqref{example11} one obtains
\begin{align}
{\vev{ \widetilde \Psi_{X}^{(w)},\widetilde \Psi_{X'}^{(w)}}}   \sim
\delta(p-p')  \int_0^1 du\, u^{2(w-s_2)-1}\bar u^{2s_2-1}\,
{}_2F_1\left(\genfrac{}{}{0pt}{}{s_2-\omega-ix,s_2-\omega+ix}{2s_2}\Big|-\frac{\bar u}{u}\right)
\notag\\
\quad\times
{}_2F_1\left(\genfrac{}{}{0pt}{}{s_2-\omega-ix',s_2-\omega+ix'}{2s_2}\Big|-\frac{\bar u}{u}\right)\,.
\end{align}
If $x=x'$ the $u$-integral is finite only if $s_2-w-ix=-n$ and $w-s_2>n \geq 0$. 
In this case the hypergeometric function reduces to a polynomial of degree $n$ in $(\bar u/u)$ and the condition  $w-s_2>n \geq
0$ guarantees that the integrand has no pole at $u=0$. 

The spin chains which are relevant for the heavy hadron phenomenology
correspond to the $s=1$, $N=2$  closed  spin chain \cite{Braun:2014npa}
and  $ s_1=\mfrac32$, $s_2=1$,  $w=\mfrac32$ open spin chain~\cite{Braun:2015pha}.

\section{Light-to-heavy reduction}\label{sect:reduction}

Physics behind the construction of the HQET is that one restricts oneself to the situations where the
heavy quark interacts with light particles (quarks and gluons) with momenta that are much smaller than the
quark mass $m_Q$ . In this case the heavy quark becomes almost stationary in its rest frame,
with its wave function oscillating rapidly with time $\Psi(t) \sim e^{-im_Qt}$ so that the generator of translations
reduces to $S_-^{(h)} \sim -im_Q$. Physical intuition suggests that this limit can be studied starting from usual $sl(2)$
algebra and rescaling the symmetry generators acting on the heavy quark
\begin{align}\label{rescaling}
     S_-^{(h)}  \to \lambda S_-^{(h)} \,,\qquad  S_+^{(h)} \to \lambda^{-1} S_+^{(h)}\,, \qquad \lambda \to \infty\,.
\end{align}
In this Section we investigate this possibility.

Consider a system consisting of one heavy quark $(h)$ and one light quark $(q)$. The two-particle generators are
\begin{align}
           S_+^{(qh)} &\equiv   S_+^{(q)} +  S_+^{(h)} \mapsto   S_+^{(q)} +  \lambda^{-1}  S_+^{(h)} =  S_+^{(q)} + \mathcal{O}(\lambda^{-1})\,,
\notag\\
           S_-^{(qh)} &\equiv    S_-^{(q)} + S_-^{(h)} \mapsto  S_-^{(q)} + \lambda  S_-^{(h)}   =  \lambda S_-^{(h)} + \mathcal{O}(1)\,,
\notag\\
           S_0^{(qh)} &\equiv   S_0^{(q)} +  S_0^{(h)} = \mathcal{O}(1)\,.
\end{align}
The two-particle quadratic Casimir operator becomes
\begin{align}
     S_{qh}^2 =    S_+^{(qh)}  S_-^{(qh)} +  S_0^{(qh)}({S_0^{(qh)}}-1) \mapsto \lambda  S_+^{(q)}  S_-^{(h)} + \mathcal{O}(1)
\end{align}
and the two-particle Hamiltonian \eqref{Hik} simplifies to
\begin{align}
  \mathcal{H}_{qh} =  2 \Big[\psi(J_{qh})-\psi(1)\Big]
\mapsto \ln\Big(\lambda S_+^{(q)}  S_-^{(h)} \Big) + \mathcal{O}(\lambda^{-1})\,.
\label{H2-reduction}
\end{align}
Since the ``heavy'' and ``light'' generators act on different spaces we can write, omitting the inessential constant
\begin{align}
    \mathcal{H}_{qh}  \mapsto  {\ln\Big(i\mu S_+^{(q)}\Big)} +  \ln\Big(-i\mu^{-1} \lambda S_-^{(h)}\Big)
\label{facH}
\end{align}
where $\mu$ is an arbitrary parameter with dimension of mass. Thus the heavy and light degrees of freedom decouple
from one another which is the statement of factorization,  with $\mu$ being the factorization scale.
In HQET only light degrees of freedom remain so that the second part of the Hamiltonian in \eqref{facH} is dropped
and the remaining part
\begin{align}
      \mathcal{H}_{qh}^{\rm HQET} =  \ln\Big(i\mu S_+^{(q)}\Big)
\end{align}
coincides with the heavy-light Hamiltonian~\eqref{Hqh} (up to the scheme-dependent constant) found in
~\cite{Lange:2003ff,Braun:2003wx,Knodlseder:2011gc} by explicit calculation of one-loop diagrams.

Next consider the eigenfunctions. Let $z_q$ and $z_h$ be the positions of light and heavy quarks, respectively.
For light quark systems, eigenfunctions of the Hamiltonian are usually sought on the space of polynomials that
can be mapped to local composite operators, see. e.g.~\cite{Braun:2003rp}. For the two-particle Hamiltonian
in \eqref{H2-reduction} the eigenfunctions on this space are well known
\begin{align}
    \Psi_{n,k}(z_q,z_h) = (S_+^{(qh)})^k (z_q-z_h)^n\,, \qquad   S_{qh}^2  \Psi_{n,k} = (n+2)(n+1)  \Psi_{n,k},
\end{align}
where $k$ is a non-negative integer. For our purposes we need eigenfunctions analytic in the
lower half of the complex plane that can be constructed as follows:
\begin{align}
\Psi_n^{(\eta)}(z_q,z_h) &= e^{-\frac i\eta S_+}(z_q-z_h)^n =  \sum_k \frac{1}{k!}\left(-\frac i\eta  S_+ \right)^k(z_q-z_h)^n
\sim \frac{(z_q-z_h)^n}{(z_q-i \eta)^{n+2}(z_h-i\eta)^{n+2}},
\label{eta-states}
\end{align}
where $\eta$ is a parameter. For $\text{Re}(\eta)>0$ they have a finite norm with respect to the scalar product~\eqref{scpN}.

The scaling in \eqref{rescaling} corresponds to $z_h \to \lambda^{-1} z_h $ so that $z_h \ll z_q$ and also
$n = \mathcal{O}(\lambda^{1/2})$.
Extracting the leading behavior at $\lambda\to\infty$ one breaks the conformal symmetry so that the states
\eqref{eta-states} with different $\eta$ are no longer degenerate and the system ``chooses'' a
particular solution that satisfies the residual symmetry to the special conformal transformations,
\begin{align}
  \Psi_n^{(\eta)}(z_q,z_h)  \mapsto \Psi_{(s)} (z_q) \,, \qquad S_+^{(q)}  \Psi_{(s)} (z_q) = i s\,  \Psi_{(s)} (z_q)\,.
\end{align}
Using
\begin{align}
 S_+^{(qh)} \Psi_n^{(\eta)}(z_q,z_h) &
= \biggl\{ \frac{z_q^2}{z_q-i\eta}\left[ n \frac{z_h-i\eta}{z_q-z_h} - 2 \frac{i\eta}{z_q} \right] + (z_h\leftrightarrow z_q) \biggr\} \Psi_n^{(\eta)}(z_q,z_h)
\end{align}
it is easy to convince oneself that the expression in the braces reduces 
in the $\lambda\to\infty$ limit and $n = \mathcal{O}(\lambda^{1/2})$ to a finite constant, $\{\ldots\} \mapsto i n \eta \equiv i s $,
if and only if $\eta = \mathcal{O}(\lambda^{-1/2})$ so that $z_h \ll \eta \ll z_q$.
The eigenfunction then becomes
\begin{align}
  \Psi_{(s)} (z_q) &=
\Psi_n^{(\eta)}(z_q,z_h)\big|_{\lambda\to\infty} \simeq \frac{1}{(i\eta)^{n+2} z_q^2} \left(1+\frac{i\eta}{z_q}\right)^n
(1\!+\!\mathcal{O}(\eta))=\frac{1}{(i\eta)^{n+2}}
\frac1{z_q^2} e^{is/z_q}\Big(1\!+\!\mathcal{O}(\lambda^{-1/2})\Big),
\label{Hwavefunction}
\end{align}
reproducing the result in \eqref{eigenH}~\cite{Braun:2014owa} up to a different normalization.

It is easy to convince oneself that the same reduction procedure applies to the conserved charges for both closed and open spin chains.
For this discussion it is convenient to enumerate sites of the chain by $i=0,1,\ldots, N$ and associate the ``heavy quark''
with the site $i=0$, $z_0\mapsto\lambda^{-1} z_h$.

For a closed spin chain one can start from the monodromy matrix \eqref{monodromy-closed} for the system of
$N$ light and one heavy quark,
\begin{align}
  T_{N+1}(u) = \left\{u + i\begin{pmatrix}S^{(h)}_0 & S^{(h)}_-\\ S^{(h)}_+ & -S^{(h)}_0 \end{pmatrix}\right\}\,
\begin{pmatrix}A_N(u) & B_N(u)\\C_N(u) & D_N(u) \end{pmatrix},
\end{align}
so that rescaling the heavy quark generators  \eqref{rescaling} one obtains the transfer matrix \eqref{t-closed}
\begin{align}\label{t-to-C}
   t_{N+1}(u) \mapsto \lambda i S_-^{(h)} C_N(u) + \mathcal{O}(\lambda^0)
\end{align}
assuming $u  \ll \lambda$. Dismissing the prefactor $ \lambda i S_-^{(h)}$ that acts on the heavy degrees of freedom we
are thus left with a family of conserved charges $C_N$ acting on the light quarks. For the simplest case of
leading-twist distribution amplitudes of heavy baryons considered in \cite{Braun:2014npa} one obtains
\begin{align}
  C_2(u) = u \mathbb{Q}_1 + \mathbb{Q}_2\,, \qquad  \mathbb{Q}_1  = i (S_+^{(1)}+ S_+^{(2)})\,, \qquad  \mathbb{Q}_2 = S_0^{(1)}S_+^{(2)} - S_0^{(2)}S_+^{(1)}.
\end{align}

Exactly in the same way one finds that the transfer matrix for the open spin chain $\mathbf{t}_{N+1}^{(w_0,w_N)}(u)$,  see
Eq.~\eqref{TMLight},
takes the form
\begin{align}
\mathbf{t}_{N+1}^{(w_0,w_N)}(u)\Big|_{z_0\to  \frac{z_h}\lambda} &\underset{\lambda\to\infty}{\longmapsto}
\lambda 2i\Big(u-\frac i2\Big) S_-^{(h)} \mathbb{C}^{(w_N)}_{N}(u) + \mathcal{O}(\lambda^0).
\end{align}
For the simplest case $N=2$  we get
\begin{align}
\mathbb{C}_2(u) = 2i\,\left(u+\frac i 2\right)\left(u^2 \mathbbm{Q}_1+\mathbbm{Q}_2\right),
\end{align}
where
\begin{align}\label{Q1Q2}
 \mathbbm{Q}_1&=S_+^{(1)}+ S_+^{(2)},   \notag\\
\mathbbm{Q}_2&=[\omega_2^2-j_2(j_2-1)] S_+^{(1)} - S^{(1)}_+ \Big(S^{(1)}_+S^{(2)}_- + S^{(1)}_0 S^{(2)}_0\Big)
-S^{(1)}_0\Big(S^{(2)}_0 S^{(1)}_+ - S^{(1)}_0 S^{(2)}_+\Big).
\end{align}
For the twist-four heavy-light operators considered in~\cite{Braun:2017liq} there is additional complication.
In this situation  several operators exist which mix together by the RG equations. i.e. the relevant Hamiltonians have
matrix structure. This case is considered in the Appendix.

Thus we see that conserved charges and Hamiltonians of the spin chain models
which describe the scale dependence of the light quark-gluon operators
in the ``heavy-quark'' limit $z_h\to z_h/\lambda$, $z_h\partial_h\sim O(\lambda^0)$,  $\lambda\to\infty$
go over to the conserved charges and Hamiltonians of the spin chains that arise in studies of the scale
dependence of heavy-light operators. This observation alone is, however, not sufficient
to guarantee equivalence of the spectra of these models; it is only true if
the correspondence can be extended to their eigenfunctions.
As shown above, this correspondence indeed holds for the eigenfunctions of the two-particle
Hamiltonian, Eq.~\eqref{Hwavefunction}, i.e. for $N=1$. The general case $N>1$ (we remind that $N$ refers
to the number of the remaining light degrees of freedom) is more complicated.
From the general concept behind effective field theories it is natural to expect that such correspondence
exists for the lowest part of the spectrum and, indeed, we are able to verify its existence for all cases
of physical relevance that have been considered so far. Whether this conclusion can be extended beyond
these examples, is not obvious.

Consider closed spin chains first. To this end we start from the ``light'' chain with $N+1$ sites, $i = 0,1,\ldots,N$.
The corresponding eigenstates diagonalize the transfer matrix
$t_{N+1}(u)$ which is a polynomial  of degree $N+1$ in $u$, $t_{N+1}(u) =2 u^{N+1} + \sum_{k=2}^{N+1} q_k u^{N+1-k}$.
The eigenstates can be labeled by eigenvalues of the conserved charges $q_k$, or, alternatively,
by roots of the transfer matrix $u_k$, $k=1,\ldots,N+1$,  $t_{N+1}(u_k)=0$.
Note that $\sum_k u_k=0$, since the subleading in $u$ term $\sim u^N $ in the transfer matrix is absent.

We found that in the limit~\eqref{rescaling} the transfer matrix for $u\ll \lambda$ takes the form $t_{N+1}(u)\sim \lambda
C_{N}(u)$~\eqref{t-to-C}, where $C_{N}(u)$ is a polynomial of degree $N-1$. This means that some of the roots of the transfer
matrix must become large in this limit, and the simplest way how Eq.~\eqref{t-to-C} may hold is when a pair of roots move to
infinity with opposite sign, say, $u_{N+1} \sim - u_{N}  \sim \sqrt{\lambda}$. This asymptotic behavior corresponds to the
situation when all conserved charges are large and of the same order, $q_k\sim \lambda$. Eigenstates for which the conserved
charges satisfy this relation  are close to the lower boundary of the energy spectrum, see
e.g.~\cite{Belitsky:2006en,Beccaria:2007uj}.

Our conjecture is that for such eigenstates the correspondence between the large-spin limit $n\sim \sqrt{\lambda} \to \infty$
of the ``light'' spin chain and the ``heavy-light'' chain holds, of the form
\begin{align}\label{conjecture-1}
\Psi^{(\eta\sim s/\sqrt{\lambda})}_{u_1,\ldots,u_{N+1}}(z_1,\ldots, z_{N}, z_0=z_h/{\lambda})
& \underset{\lambda\to\infty}{\longmapsto}
\Psi^{\rm heavy}_{s,u_1,\ldots,u_{N-1}}(z_1,\ldots,z_N).
\end{align}
where $\Psi^{\rm heavy}_{s,u_1,\ldots,u_{N-1}}(z_1,\ldots,z_N)$ is the eigenstate of the ``heavy-light'' spin chain.

This correspondence can be illustrated by the following example. The lowest-energy eigenfunctions for the closed spin
chain with three sites (spin $s=1$) correspond to the solutions  with $q_3=0$ and
are known explicitly for even $n$~\cite{Braun:1999te}
\begin{align}
\Psi_{n,q_3=0}(z_0,z_1,z_2)=\frac{z_{01}^{n+3}+z_{12}^{n+3}+z_{20}^{n+3}}{z_{01} z_{12} z_{20}}\,,\qquad z_{ik} = z_i-z_k\,.
\end{align}
Mapping these polynomial solutions to analytic functions of the coordinates in the
lower half-plane, cf.~\eqref{eta-states}, one obtains
\begin{align}
\Psi^{(\eta)}_{n,q_3=0}(z_0,z_1,z_2)\sim \frac{\left(\frac{1}{z_0-i\eta}-\frac{1}{z_1-i\eta}\right)^{n+3}
\!\!+\left(\frac{1}{z_1-i\eta}-\frac{1}{z_2-i\eta}\right)^{n+3}
\!\!+\left(\frac{1}{z_2-i\eta}-\frac{1}{z_0-i\eta}\right)^{n+3}}{z_{01} z_{12} z_{20}}\,.
\end{align}
In the limit $z_0 \sim 1/\lambda$, $\eta \sim 1/\sqrt{\lambda}$, $n\sim \sqrt{\lambda}$, $s = \eta n$,
this expression goes over to
\begin{align}
\Psi_{n,q_3=0}(z_0,z_1,z_2)\underset{\lambda\to\infty}{\longmapsto} \frac{1}{z_1 z_2 z_{12}} \left(e^{is/z_1}-e^{is/z_2}\right),
\end{align}
which coincides with the solution $\phi_{s,x=0}$ obtained in \cite{Braun:2014npa}.
For higher-lying states, expressions for the eigenfunctions of three light quarks are not known in explicit form
so that verifying this correspondence is less straightforward. 
However, it is possible to check that the energies expressed in terms of the roots $u_k$ indeed coincide in this limit.

The corresponding construction for open spin chains is analogous.
In this case we also can label the eigenfunctions of the original ``light'' model
with $N+1$ sites by the roots of the transfer matrix, which come in pairs $\pm u_k$, $k=1,\ldots, N+1$.
We consider the limit when one root is large, $u_{N+1} \sim \sqrt{\lambda}$, and all
others are finite, $O(\lambda^0)$. Similar to closed chains such a hierarchy corresponds
to the situation when all charges are large and of the same order.
We notice also that one of the roots in this limit is purely kinematical, $u_{N}=i/2$.
Thus the limiting eigenfunction is labeled by the remaining $N-1$
roots of the transfer matrix and depends also on the parameter $\eta$ related to the residual
symmetry transformation.
We, therefore, expect that the conjecture \eqref{conjecture-1} may hold for the lowest
eigenstates of open spin chains as well.

As an example, consider the open spin chain with three sites that arises in the description of
the evolution of quark-antiquark-gluon operators of twist three in the large
$N_c$ limit~\cite{Braun:1998id,Belitsky:1999ru,Derkachov:1999ze}.
This is a model with spins  ${(s_0,s_1,s_2)=(1,\frac32,1)}$ and impurities $(w_0,w_1,w_2)=(\frac12,0,\frac32)$.
Its eigenstates satisfy the equation  $\mathbbm{Q}_4\Psi_n=q_S(n) \Psi_n$ where
\begin{align}\label{Q-4-charge}
\mathbbm{Q}_4=\{S^2_{01}, S^2_{12}\}-2w_2S^2_{01} -2 w_0 S^2_{12}\,.
\end{align}
Here $\{.,.\}$ stands for the anticommutator and $S^2_{ik}$
are the two-particle quadratic Casimir operators.

The lowest-energy eigenstate corresponds to the eigenvalue $q_S(n)=n(n+6)+\frac{75}8$~\cite{Derkachov:1999ze}
and can be found in explicit form,%
\footnote{This is a new result.}
\begin{align}
\Psi_n(z_0,z_1,z_2)
&=\frac1{z_{01}z_{12}^2}\biggl[\frac{n+3}{n+4}z_{10}^{n+3}+z_{20}^{n+3}+\frac1{n+4}\left(\frac{z_{20}^{n+4}-z_{21}^{n+4}}{z_{01}}\right)
+
\frac{2}{z_{12}}\left(\frac{z_{20}^{n+4}}{n+4}-\frac{z_{10}^{n+4}}{n+5}\right)
\notag\\
&\hspace*{0.7cm}{}+\frac2{(n+4)(n+5)}\frac{z_{20}^{n+5}-z_{21}^{n+5}}{z_{01}z_{12}}\biggr].
\end{align}
Mapping this polynomial solution to an analytic function in the lower half-plane
\begin{align}
\Psi_n\to \Psi_n^{{(\eta)}} = e^{-\frac i\eta S_+}\Psi_n(z_0,z_1,z_2)
\sim \prod_{k=0}^2\frac{1}{(z_k-i\eta)^{2s_k}}
\Psi_n
\left(\frac{z_0}{z_0-i\eta},\frac{z_1}{z_1-i\eta},\frac{z_2}{z_2-i\eta}\right)
\end{align}
and taking the limit $\lambda\to\infty$ with
$z_0 \sim 1/\lambda$, $\eta \sim 1/\sqrt{\lambda}$, $n\sim \sqrt{\lambda}$, $s = \eta n$, one easily finds
\begin{align}
 \Psi_n^{{(\eta)}} (z_0,z_1,z_2)\underset{\lambda\to\infty}{\longmapsto}
 \frac1{z_1 z_{12}^2}\left[ e^{is/z_1}+e^{is/z_2}+\frac{2 z_1 z_2}{is z_{12}}\left( e^{is/z_1} - e^{is/z_2}\right)\right].
\end{align}
This expression coincides with the discrete state solution $Y_{s}^{(0)}(z_1,z_2)$ for the heavy quark -- light antiquark -- gluon
RG equation
found in~\cite{Braun:2015pha}.

A word of caution has to be added. Although the given examples support the conjecture in Eq.~\eqref{conjecture-1},
we do not claim that there exists a one-to-one correspondence between the eigenfunctions of generic
light models in the large spin limit and heavy-light models. A counterexample can easily be found.
Indeed, in the same open chain model considered above one can try to make the particle with $i=2$ heavy instead of $i=0$,
i.e. consider the limit  $z_2=z_h/\lambda \to 0 $ instead of $z_0\to 0$. It turns out that the eigenfunction
in this limit becomes non-normalizable. Thus the lowest energy discrete state of the ``light'' chain
disappears from the spectrum and is not present in the corresponding ``heavy-light'' chain.
A detailed study of this phenomenon goes beyond the tasks of this work.

Another issue is that beyond leading order the identification of the effective theory acting on light degrees of freedom
as HQET is expected to break down because of contributions of hard-collinear gluon emission and a more general, e.g.
soft-collinear effective theory, may arise. This is one more topic for future study.

\section{Summary}
Evolution equations for many physically relevant heavy-light operators in QCD turn out to be integrable in the multi-color limit.
The novelty of these systems is that the corresponding conserved charges are generated by the off-diagonal element of the
monodromy matrix $C(u)$ instead of the trace of monodromy matrix as it happens for the light quark-gluon operators. These
evolution equations can be solved with the help of QISM, resulting in a better understanding of the
structure of the $B$-meson distribution amplitudes of leading and subleading twists~\cite{Braun:2015pha,Braun:2017liq}.
The aim of this paper is to present a more detailed mathematical treatment of such spin chain models in QISM
formalism and explain details of the derivation omitted in~\cite{Braun:2015pha,Braun:2017liq}. This is done in
Sects.~\ref{sect:HLchains}-\ref{sec:Eigenfunctions}.

Another aim is  to explore the observation made in~\cite{Braun:2014npa,Braun:2015pha} that a certain 
similarity exists between the spectrum of the heavy-light spin chains and ordinary $\text{SL}(2,\mathbb{R})$-invariant 
spin chains in the large spin limit~\cite{Belitsky:2006en,Beccaria:2007uj}. One finds on several examples 
that the expressions for the ground state wave functions, mass gaps, etc. in both models coincide.  
In Sec.~\ref{sect:reduction} we suggest an explicit mapping behind this correspondence.
It turns out that that sending $z_h\to 0$, where $z_h$ is a coordinate associated with the ``heavy quark'', 
the Hamiltonian and conserved charges of the ``light'' spin chain models factorize and go over to 
the Hamiltonian and conserved charges of the ``heavy-light'' models.

We further argue that such correspondence extends to the low-lying states at the eigenfunction level 
in a certain scaling limit, which means that such states can be studied 
using effective theory methods. In particular, eigenfunctions of the lowest energy states 
with large spin in the ``light'' sector can be approximated by the eigenfunctions of the conserved charge 
$\mathbb{C}(u)$, which can be constructed with the help of Sklyanin's  method of Separation of Variables (SoV). 
We have checked that this conjecture indeed works in several cases where analytic expressions for the eigenstates are known.
Alternative method for the analysis of low-lying states for the ``light'' spin chain models in the large spin limit 
is based on the the semiclassical expansion of the solution of the Baxter equation, see e.g.~\cite{Belitsky:2006en,Beccaria:2007uj}.
In this approach the wave functions of ``light'' models can be obtained as a 
convolution of the corresponding Baxter functions with the transition kernel to the SoV representation 
which is nothing but an eigenstate of the ``heavy-light'' spin chain,
inviting for an effective theory interpretation of this method.
One can hope that using QISM technique it will be possible to give this interpretation a more precise meaning.

\section*{Acknowledgments}
We are grateful to S. Derkachov and G. Korchemsky for useful discussions.
The work  was supported by the DFG grants BR 2021/7-1 (YJ), MO~1801/1-3 (AM) and by RSF 
project 
14-11-00598 (AM).



\section*{Appendix}
\addcontentsline{toc}{section}{Appendices}

\renewcommand{\theequation}{\Alph{section}.\arabic{equation}}
\renewcommand{\thetable}{\Alph{table}}
\setcounter{section}{0}
\setcounter{table}{0}

\section{Twist-four operators}
In this section we explain the method which was used in~\cite{Braun:2017liq}
to construct the conserved charges for the Hamiltonians describing the scale-dependence of higher-twist
heavy-light operators. To this end we consider a particular example of an operator doublet
\begin{align}\label{heavy-doublet}
\overrightarrow{\mathcal{Q}_h}= \begin{pmatrix}
\psi_-(z_1)f_{++}(z_2)h_v(0) \\
\psi_+(z_1)f_{+-}(z_2)h_v(0)
\end{pmatrix}.
\end{align}
The evolution kernel (Hamiltonian) for the operators $\overrightarrow{\mathcal{Q}_h}$ is given by a $2\times2$ matrix
\begin{align}
\mathbb{H}_h=\begin{pmatrix}
H_{11} & H_{12}\\
H_{21} & H_{22}
\end{pmatrix}.
\label{Hdoublet}
\end{align}
Explicit expressions for the kernels $H_{ik}$ can be found, e.g., in~\cite{Braun:2009vc,Braun:2009mi,Ji:2014eta}.
Our aim is to find the conserved charges $\mathbb{Q}_k$ that commute with the Hamiltonian $\mathbb{H}_h$.
This can be done using standard techniques starting from the spin chain with complete conformal
$\text{SO}(4,2)$ symmetry. The method described below allows one to stay with the $\text{SL}(2)$ subgroup
and achieve the same result in a more straightforward way.

To start with, consider the twist-three light quark-antiquark-gluon operator,
\begin{align}\label{Q-tw-3}
\mathcal{Q}_{qgq}=\psi_+(z_1)f_{++}(z_2) \psi_+(z_0)\,.
\end{align}
The corresponding evolution equation is  well-studied.  The evolution kernel in the large $N_c$ limit commutes with
the (three-particle) quadratic Casimir operator $\mathbb{S}^2$ and with an additional conserved charge,
$\mathbb{Q}_4$~\eqref{Q-4-charge} with impurity parameters, $w_1=w_2=\frac12$.
Note that  $\mathbb{Q}_4$ is written in terms of the two-particle quadratic Casimir operators of the
$\mathrm{SL}(2,\mathbb{R})$ group.

The trick is to use this result to construct the conserved charges for the twist-four operator,
\begin{align}\label{Q-tw-4}
\overrightarrow{\mathcal{Q}}= \begin{pmatrix}
\psi_-(z_1)f_{++}(z_2)\psi_+(z_0) \\
\psi_+(z_1)f_{+-}(z_2)\psi_+(z_0)\\
\psi_+(z_1)f_{++}(z_2)\psi_-(z_0) \end{pmatrix}\,.
\end{align}
The operators \eqref{Q-tw-3} and \eqref{Q-tw-4} do not mix under the collinear conformal $\mathrm{SL}(2,\mathbb{R})$ transformation,
but they are related by a transformation involving the full conformal group.
As a consequence, the conserved charge $\widehat{\mathbb{Q}}_4$ for the twist-four operators~\eqref{Q-tw-4}
can be found by promoting the $\mathrm{SL}(2,\mathbb{R})$ two-particle Casimir operators in Eq.~\eqref{Q-4-charge}
to the two-particle Casimir operators of the full conformal group $\text{SO}(4,2)$, $S^2_{ik} \mapsto \widehat{\mathbb{S}}^2_{ik}$
\begin{align}\label{Q-4-big}
    \widehat{\mathbb{Q}}_4=\{\widehat{\mathbb{S}}^2_{12}, \widehat{\mathbb{S}}^2_{20}\}-2w_3\widehat{\mathbb{S}}^2_{12} -2 w_1
    \widehat{\mathbb{S}}^2_{20}
    \end{align}
and projecting the $\text{SO}(4,2)$ operators onto the twist-four subspace~\eqref{Q-tw-4}.
Explicit expressions for the generators of the full conformal group in the so-called light-ray operator representation
are given in~\cite{Braun:2008ia}. The corresponding two-particle Casimir operators
can be written as~\cite{Braun:2009vc}%
\footnote{In the corresponding expressions in~\cite{Braun:2009vc}, Eq.~(5.24),  there is a misprint: 
$\mathbb{I}$ should be $\sigma_3 = \begin{pmatrix}1 & 0 \\ 0 & -1\end{pmatrix}$.}  
\begin{align}
\widehat{\mathbb{S}}^2_{12}=\begin{pmatrix}
\widehat{J}_{12}\big(\widehat{J}_{12}-1\big)&0\\
0& {S}^2_{12}
\end{pmatrix}, && \widehat{J}_{12}= -\begin{pmatrix}
-1/2& z_{21}\partial_2+2\\
z_{12}\partial_1+1& 1/2
\end{pmatrix},
\notag\\
\widehat{\mathbb{S}}^2_{20}=\begin{pmatrix}
 {S}^2_{20}&0\\
0&\widehat{J}_{20}\big(\widehat{J}_{20}-1\big)
\end{pmatrix},
 && \widehat{J}_{20}=
-\begin{pmatrix}
-1/2& z_{02}\partial_0+1\\
z_{20}\partial_2+2& 1/2
\end{pmatrix},
\end{align}
where ${S}^2_{12}, {S}^2_{20}$ are the corresponding $\mathrm{SL}(2,\mathbb{R})$ Casimir operators:
\begin{align}
 {S}^2_{12}=-\partial_1\partial_2 z_{12}^2+ \partial_1 z_{12}+\frac34, &&
 {S}^2_{20}=-\partial_0\partial_2 z_{02}^2+ \partial_0 z_{02}+\frac34\,.
\end{align}
Using these expressions in~\eqref{Q-4-big} one obtains the conserved charge $\widehat{\mathbb{Q}}_4$
as a $3\times 3$ matrix with operator entries.

Finally, we perform the ''light-to-heavy'' reduction for this charge, $z_0 \to z_h/\lambda$, $\lambda\to \infty$. The Casimir
operator $\widehat{\mathbb{S}}^2_{20}$ factorizes in this limit:
\begin{align}
 \widehat{\mathbb{S}}^2_{20}=\lambda S^{(h)}_- \widehat{\mathbb{S}}_{2}^{(q)}  + O(\lambda^0)\,,
\end{align}
where
\begin{align}
 \widehat{\mathbb{S}}_{2}^{(q)}=\begin{pmatrix}
 z_2^{-1}\partial_2 z_2^3 &0&0\\
 0&\partial_2 z_2^2& 0\\
 0&0& z_2^{-1}\partial_2 z_2^3
 \end{pmatrix},
\end{align}
and as the result the charge $\widehat{\mathbb{Q}}_4$ takes the form
\begin{align} \widehat{\mathbb{Q}}_4=\lambda S_-^{(h)}
\widehat{\mathbb{Q}}_4^{(q)}+O(\lambda^0)\,,
\end{align}
where $\widehat{\mathbb{Q}}_4^{(q)}$ is a two-particle operator acting on the remaining
light degrees of freedom,
\begin{align}
\widehat{\mathbb{Q}}_4^{(q)}=
\{\widehat{\mathbb{C}}_{12}, \widehat{\mathbb{C}}^{(q)}_{2}\} -2 w^2_1 \widehat{\mathbb{C}}_{2}^{(q)}
=   \begin{pmatrix}
\mathcal{Q}_4^{(q)} & 0\\ 0&  \mathbbm{Q}_2
    \end{pmatrix}\,.
\end{align}
Here $\mathcal{Q}_4^{(q)} $ is a $2\times 2$ matrix, which commutes with the evolution kernel \eqref{Hdoublet} for the
operator doublet~\eqref{heavy-doublet} and the remaining entry  $\mathbbm{Q}_2$ is the conserved charge~\eqref{Q1Q2}. The
conserved charges for all other twist-four heavy-light operators considered in Ref.~\cite{Braun:2017liq} can be obtained in
this manner.


\bibliography{BJM}

\providecommand{\href}[2]{#2}\begingroup\raggedright\begin{thebibliography}{10}

\bibitem{Neubert:1993mb}
M.~Neubert, \emph{{Heavy quark symmetry}},
  \href{http://dx.doi.org/10.1016/0370-1573(94)90091-4}{\emph{Phys. Rept.}
  {\bfseries 245} (1994) 259--396},
  [\href{https://arxiv.org/abs/hep-ph/9306320}{{\ttfamily hep-ph/9306320}}].

\bibitem{Lipatov:1993yb}
L.~N. Lipatov, \emph{{Asymptotic behavior of multicolor QCD at high energies in
  connection with exactly solvable spin models}}, {\emph{JETP Lett.} {\bfseries
  59} (1994) 596--599}, [\href{https://arxiv.org/abs/hep-th/9311037}{{\ttfamily
  hep-th/9311037}}].

\bibitem{Faddeev:1994zg}
L.~D. Faddeev and G.~P. Korchemsky, \emph{{High-energy QCD as a completely
  integrable model}},
  \href{http://dx.doi.org/10.1016/0370-2693(94)01363-H}{\emph{Phys. Lett.}
  {\bfseries B342} (1995) 311--322},
  [\href{https://arxiv.org/abs/hep-th/9404173}{{\ttfamily hep-th/9404173}}].

\bibitem{Braun:1998id}
V.~M. Braun, S.~E. Derkachov and A.~N. Manashov, \emph{{Integrability of three
  particle evolution equations in QCD}},
  \href{http://dx.doi.org/10.1103/PhysRevLett.81.2020}{\emph{Phys. Rev. Lett.}
  {\bfseries 81} (1998) 2020--2023},
  [\href{https://arxiv.org/abs/hep-ph/9805225}{{\ttfamily hep-ph/9805225}}].

\bibitem{Braun:1999te}
V.~M. Braun, S.~E. Derkachov, G.~P. Korchemsky and A.~N. Manashov,
  \emph{{Baryon distribution amplitudes in QCD}},
  \href{http://dx.doi.org/10.1016/S0550-3213(99)00265-5}{\emph{Nucl. Phys.}
  {\bfseries B553} (1999) 355--426},
  [\href{https://arxiv.org/abs/hep-ph/9902375}{{\ttfamily hep-ph/9902375}}].

\bibitem{Belitsky:1999bf}
A.~V. Belitsky, \emph{{Renormalization of twist - three operators and
  integrable lattice models}},
  \href{http://dx.doi.org/10.1016/S0550-3213(00)00003-1}{\emph{Nucl. Phys.}
  {\bfseries B574} (2000) 407--447},
  [\href{https://arxiv.org/abs/hep-ph/9907420}{{\ttfamily hep-ph/9907420}}].

\bibitem{Faddeev:1979gh}
L.~D. Faddeev, E.~K. Sklyanin and L.~A. Takhtajan, \emph{{The Quantum Inverse
  Problem Method. 1}}, {\emph{Theor. Math. Phys.} {\bfseries 40} (1980)
  688--706}.

\bibitem{Takhtajan:1979iv}
L.~A. Takhtajan and L.~D. Faddeev, \emph{{The Quantum method of the inverse
  problem and the Heisenberg XYZ model}}, {\emph{Russ. Math. Surveys}
  {\bfseries 34} (1979) 11--68}.

\bibitem{Kulish:1981gi}
P.~P. Kulish, N.~{\relax Yu}. Reshetikhin and E.~K. Sklyanin,
  \emph{{Yang-Baxter Equation and Representation Theory. 1.}},
  \href{http://dx.doi.org/10.1007/BF02285311}{\emph{Lett. Math. Phys.}
  {\bfseries 5} (1981) 393--403}.

\bibitem{Sklyanin:1995bm}
E.~K. Sklyanin, \emph{{Separation of variables - new trends}},
  \href{http://dx.doi.org/10.1143/PTPS.118.35}{\emph{Prog. Theor. Phys. Suppl.}
  {\bfseries 118} (1995) 35--60},
  [\href{https://arxiv.org/abs/solv-int/9504001}{{\ttfamily
  solv-int/9504001}}].

\bibitem{Braun:2000av}
V.~M. Braun, G.~P. Korchemsky and A.~N. Manashov, \emph{{Evolution of twist -
  three parton distributions in QCD beyond the large N(c) limit}},
  \href{http://dx.doi.org/10.1016/S0370-2693(00)00131-3}{\emph{Phys. Lett.}
  {\bfseries B476} (2000) 455--464},
  [\href{https://arxiv.org/abs/hep-ph/0001130}{{\ttfamily hep-ph/0001130}}].

\bibitem{DeVega:2001pu}
H.~J. De~Vega and L.~N. Lipatov, \emph{{Interaction of reggeized gluons in the
  Baxter-Sklyanin representation}},
  \href{http://dx.doi.org/10.1103/PhysRevD.64.114019}{\emph{Phys. Rev.}
  {\bfseries D64} (2001) 114019},
  [\href{https://arxiv.org/abs/hep-ph/0107225}{{\ttfamily hep-ph/0107225}}].

\bibitem{Derkachov:2001yn}
S.~E. Derkachov, G.~P. Korchemsky and A.~N. Manashov, \emph{{Noncompact
  Heisenberg spin magnets from high-energy QCD: 1. Baxter Q operator and
  separation of variables}},
  \href{http://dx.doi.org/10.1016/S0550-3213(01)00457-6}{\emph{Nucl. Phys.}
  {\bfseries B617} (2001) 375--440},
  [\href{https://arxiv.org/abs/hep-th/0107193}{{\ttfamily hep-th/0107193}}].

\bibitem{Braun:2009vc}
V.~M. Braun, A.~N. Manashov and J.~Rohrwild, \emph{{Renormalization of
  Twist-Four Operators in QCD}},
  \href{http://dx.doi.org/10.1016/j.nuclphysb.2009.10.005}{\emph{Nucl. Phys.}
  {\bfseries B826} (2010) 235--293},
  [\href{https://arxiv.org/abs/0908.1684}{{\ttfamily 0908.1684}}].

\bibitem{Minahan:2002ve}
J.~A. Minahan and K.~Zarembo, \emph{{The Bethe ansatz for N=4
  superYang-Mills}},
  \href{http://dx.doi.org/10.1088/1126-6708/2003/03/013}{\emph{JHEP} {\bfseries
  03} (2003) 013}, [\href{https://arxiv.org/abs/hep-th/0212208}{{\ttfamily
  hep-th/0212208}}].

\bibitem{Beisert:2010jr}
N.~Beisert et~al., \emph{{Review of AdS/CFT Integrability: An Overview}},
  \href{http://dx.doi.org/10.1007/s11005-011-0529-2}{\emph{Lett. Math. Phys.}
  {\bfseries 99} (2012) 3--32},
  [\href{https://arxiv.org/abs/1012.3982}{{\ttfamily 1012.3982}}].

\bibitem{Braun:2014npa}
V.~M. Braun, S.~E. Derkachov and A.~N. Manashov, \emph{{Integrability of the
  evolution equations for heavy–light baryon distribution amplitudes}},
  \href{http://dx.doi.org/10.1016/j.physletb.2014.09.062}{\emph{Phys. Lett.}
  {\bfseries B738} (2014) 334--340},
  [\href{https://arxiv.org/abs/1406.0664}{{\ttfamily 1406.0664}}].

\bibitem{Braun:2015pha}
V.~M. Braun, A.~N. Manashov and N.~Offen, \emph{{Evolution equation for the
  higher-twist B-meson distribution amplitude}},
  \href{http://dx.doi.org/10.1103/PhysRevD.92.074044}{\emph{Phys. Rev.}
  {\bfseries D92} (2015) 074044},
  [\href{https://arxiv.org/abs/1507.03445}{{\ttfamily 1507.03445}}].

\bibitem{Braun:2017liq}
V.~M. Braun, Y.~Ji and A.~N. Manashov, \emph{{Higher-twist B-meson Distribution
  Amplitudes in HQET}},  \href{https://arxiv.org/abs/1703.02446}{{\ttfamily
  1703.02446}}.

\bibitem{Lipatov:2009nt}
L.~N. Lipatov, \emph{{Integrability of scattering amplitudes in N=4 SUSY}},
  \href{http://dx.doi.org/10.1088/1751-8113/42/30/304020}{\emph{J. Phys.}
  {\bfseries A42} (2009) 304020},
  [\href{https://arxiv.org/abs/0902.1444}{{\ttfamily 0902.1444}}].

\bibitem{Basso:2010in}
B.~Basso, \emph{{Exciting the GKP string at any coupling}},
  \href{http://dx.doi.org/10.1016/j.nuclphysb.2011.12.010}{\emph{Nucl. Phys.}
  {\bfseries B857} (2012) 254--334},
  [\href{https://arxiv.org/abs/1010.5237}{{\ttfamily 1010.5237}}].

\bibitem{Bartels:2011nz}
J.~Bartels, L.~N. Lipatov and A.~Prygarin, \emph{{Integrable spin chains and
  scattering amplitudes}},
  \href{http://dx.doi.org/10.1088/1751-8113/44/45/454013}{\emph{J. Phys.}
  {\bfseries A44} (2011) 454013},
  [\href{https://arxiv.org/abs/1104.0816}{{\ttfamily 1104.0816}}].

\bibitem{Belitsky:2011nn}
A.~V. Belitsky, \emph{{OPE for null Wilson loops and open spin chains}},
  \href{http://dx.doi.org/10.1016/j.physletb.2012.02.027}{\emph{Phys. Lett.}
  {\bfseries B709} (2012) 280--284},
  [\href{https://arxiv.org/abs/1110.1063}{{\ttfamily 1110.1063}}].

\bibitem{Korchemsky:1991zp}
G.~P. Korchemsky and A.~V. Radyushkin, \emph{{Infrared factorization, Wilson
  lines and the heavy quark limit}},
  \href{http://dx.doi.org/10.1016/0370-2693(92)90405-S}{\emph{Phys. Lett.}
  {\bfseries B279} (1992) 359--366},
  [\href{https://arxiv.org/abs/hep-ph/9203222}{{\ttfamily hep-ph/9203222}}].

\bibitem{Lange:2003ff}
B.~O. Lange and M.~Neubert, \emph{{Renormalization group evolution of the B
  meson light cone distribution amplitude}},
  \href{http://dx.doi.org/10.1103/PhysRevLett.91.102001}{\emph{Phys. Rev.
  Lett.} {\bfseries 91} (2003) 102001},
  [\href{https://arxiv.org/abs/hep-ph/0303082}{{\ttfamily hep-ph/0303082}}].

\bibitem{Braun:2003wx}
V.~M. Braun, D.~{\relax Yu}. Ivanov and G.~P. Korchemsky, \emph{{The B meson
  distribution amplitude in QCD}},
  \href{http://dx.doi.org/10.1103/PhysRevD.69.034014}{\emph{Phys. Rev.}
  {\bfseries D69} (2004) 034014},
  [\href{https://arxiv.org/abs/hep-ph/0309330}{{\ttfamily hep-ph/0309330}}].

\bibitem{Knodlseder:2011gc}
M.~Knodlseder and N.~Offen, \emph{{Renormalisation of heavy-light light ray
  operators}}, \href{http://dx.doi.org/10.1007/JHEP10(2011)069}{\emph{JHEP}
  {\bfseries 10} (2011) 069},
  [\href{https://arxiv.org/abs/1105.4569}{{\ttfamily 1105.4569}}].

\bibitem{Braun:2014owa}
V.~M. Braun and A.~N. Manashov, \emph{{Conformal symmetry of the Lange-Neubert
  evolution equation}},
  \href{http://dx.doi.org/10.1016/j.physletb.2014.02.051}{\emph{Phys. Lett.}
  {\bfseries B731} (2014) 316--319},
  [\href{https://arxiv.org/abs/1402.5822}{{\ttfamily 1402.5822}}].

\bibitem{Bukhvostov:1985rn}
A.~P. Bukhvostov, G.~V. Frolov, L.~N. Lipatov and E.~A. Kuraev,
  \emph{{Evolution Equations for Quasi-Partonic Operators}},
  \href{http://dx.doi.org/10.1016/0550-3213(85)90628-5}{\emph{Nucl. Phys.}
  {\bfseries B258} (1985) 601--646}.

\bibitem{MR3469700}
I.~M. Gelfand, M.~I. Graev and N.~Y. Vilenkin, \emph{Generalized functions.
  {V}ol. 5}.
\newblock AMS Chelsea Publishing, Providence, RI, 2016.

\bibitem{Ball:2008fw}
P.~Ball, V.~M. Braun and E.~Gardi, \emph{{Distribution Amplitudes of the
  Lambda(b) Baryon in QCD}},
  \href{http://dx.doi.org/10.1016/j.physletb.2008.06.004}{\emph{Phys. Lett.}
  {\bfseries B665} (2008) 197--204},
  [\href{https://arxiv.org/abs/0804.2424}{{\ttfamily 0804.2424}}].

\bibitem{Wang:2011uv}
W.~Wang, \emph{{Factorization of Heavy-to-Light Baryonic Transitions in SCET}},
  \href{http://dx.doi.org/10.1016/j.physletb.2012.01.036}{\emph{Phys. Lett.}
  {\bfseries B708} (2012) 119--126},
  [\href{https://arxiv.org/abs/1112.0237}{{\ttfamily 1112.0237}}].

\bibitem{Ali:2012pn}
A.~Ali, C.~Hambrock, A.~{\relax Ya}. Parkhomenko and W.~Wang, \emph{{Light-Cone
  Distribution Amplitudes of the Ground State Bottom Baryons in HQET}},
  \href{http://dx.doi.org/10.1140/epjc/s10052-013-2302-4}{\emph{Eur. Phys. J.}
  {\bfseries C73} (2013) 2302},
  [\href{https://arxiv.org/abs/1212.3280}{{\ttfamily 1212.3280}}].

\bibitem{Sklyanin:1991ss}
E.~K. Sklyanin, \emph{{Quantum inverse scattering method. Selected topics}},
  \href{https://arxiv.org/abs/hep-th/9211111}{{\ttfamily hep-th/9211111}}.

\bibitem{Derkachov:2002tf}
S.~E. Derkachov, G.~P. Korchemsky and A.~N. Manashov, \emph{{Separation of
  variables for the quantum SL(2,R) spin chain}},
  \href{http://dx.doi.org/10.1088/1126-6708/2003/07/047}{\emph{JHEP} {\bfseries
  07} (2003) 047}, [\href{https://arxiv.org/abs/hep-th/0210216}{{\ttfamily
  hep-th/0210216}}].

\bibitem{Derkachov:2016dhc}
S.~E. Derkachov and A.~N. Manashov, \emph{{Spin Chains and Gustafson's
  Integrals}},  \href{https://arxiv.org/abs/1611.09593}{{\ttfamily
  1611.09593}}.

\bibitem{Derkachov:1999pz}
S.~E. Derkachov, \emph{{Baxter's Q-operator for the homogeneous XXX spin
  chain}}, \href{http://dx.doi.org/10.1088/0305-4470/32/28/309}{\emph{J. Phys.}
  {\bfseries A32} (1999) 5299--5316},
  [\href{https://arxiv.org/abs/solv-int/9902015}{{\ttfamily
  solv-int/9902015}}].

\bibitem{Belitsky:2014rba}
A.~V. Belitsky, S.~E. Derkachov and A.~N. Manashov, \emph{{Quantum mechanics of
  null polygonal Wilson loops}},
  \href{http://dx.doi.org/10.1016/j.nuclphysb.2014.03.007}{\emph{Nucl. Phys.}
  {\bfseries B882} (2014) 303--351},
  [\href{https://arxiv.org/abs/1401.7307}{{\ttfamily 1401.7307}}].

\bibitem{Derkachov:2005hw}
S.~E. Derkachov, \emph{{Factorization of the R-matrix. I.}},
  \href{https://arxiv.org/abs/math/0503396}{{\ttfamily math/0503396}}.

\bibitem{Derkachov:2005fg}
S.~E. Derkachov, \emph{{Factorization of R-matrix and Baxter's Q-operator}},
  \href{http://dx.doi.org/10.1007/s10958-008-9005-7}{\emph{J. Math. Sci.}
  {\bfseries 151} (2008) 2848--2858},
  [\href{https://arxiv.org/abs/math/0507252}{{\ttfamily math/0507252}}].

\bibitem{Sklyanin:1988yz}
E.~K. Sklyanin, \emph{{Boundary Conditions for Integrable Quantum Systems}},
  \href{http://dx.doi.org/10.1088/0305-4470/21/10/015}{\emph{J. Phys.}
  {\bfseries A21} (1988) 2375--289}.

\bibitem{Derkachov:2003qb}
S.~E. Derkachov, G.~P. Korchemsky and A.~N. Manashov, \emph{{Baxter Q operator
  and separation of variables for the open SL(2,R) spin chain}},
  \href{http://dx.doi.org/10.1088/1126-6708/2003/10/053}{\emph{JHEP} {\bfseries
  10} (2003) 053}, [\href{https://arxiv.org/abs/hep-th/0309144}{{\ttfamily
  hep-th/0309144}}].

\bibitem{Derkachov:1999ze}
S.~E. Derkachov, G.~P. Korchemsky and A.~N. Manashov, \emph{{Evolution
  equations for quark gluon distributions in multicolor QCD and open spin
  chains}}, \href{http://dx.doi.org/10.1016/S0550-3213(99)00702-6}{\emph{Nucl.
  Phys.} {\bfseries B566} (2000) 203--251},
  [\href{https://arxiv.org/abs/hep-ph/9909539}{{\ttfamily hep-ph/9909539}}].

\bibitem{Braun:2003rp}
V.~M. Braun, G.~P. Korchemsky and D.~Mueller, \emph{{The Uses of conformal
  symmetry in QCD}},
  \href{http://dx.doi.org/10.1016/S0146-6410(03)90004-4}{\emph{Prog. Part.
  Nucl. Phys.} {\bfseries 51} (2003) 311--398},
  [\href{https://arxiv.org/abs/hep-ph/0306057}{{\ttfamily hep-ph/0306057}}].

\bibitem{Belitsky:2006en}
A.~V. Belitsky, A.~S. Gorsky and G.~P. Korchemsky, \emph{{Logarithmic scaling
  in gauge/string correspondence}},
  \href{http://dx.doi.org/10.1016/j.nuclphysb.2006.04.030}{\emph{Nucl. Phys.}
  {\bfseries B748} (2006) 24--59},
  [\href{https://arxiv.org/abs/hep-th/0601112}{{\ttfamily hep-th/0601112}}].

\bibitem{Beccaria:2007uj}
M.~Beccaria and F.~Catino, \emph{{Large spin expansion of the long-range Baxter
  equation in the sl(2) sector of N=4 SYM}},
  \href{http://dx.doi.org/10.1088/1126-6708/2008/01/067}{\emph{JHEP} {\bfseries
  01} (2008) 067}, [\href{https://arxiv.org/abs/0710.1991}{{\ttfamily
  0710.1991}}].

\bibitem{Belitsky:1999ru}
A.~V. Belitsky, \emph{{Integrability and WKB solution of twist - three
  evolution equations}},
  \href{http://dx.doi.org/10.1016/S0550-3213(99)00402-2}{\emph{Nucl. Phys.}
  {\bfseries B558} (1999) 259--284},
  [\href{https://arxiv.org/abs/hep-ph/9903512}{{\ttfamily hep-ph/9903512}}].

\bibitem{Braun:2009mi}
V.~M. Braun, A.~N. Manashov and B.~Pirnay, \emph{{Scale dependence of
  twist-three contributions to single spin asymmetries}},
  \href{http://dx.doi.org/10.1103/PhysRevD.80.114002,
  10.1103/PhysRevD.86.119902}{\emph{Phys. Rev.} {\bfseries D80} (2009) 114002},
  [\href{https://arxiv.org/abs/0909.3410}{{\ttfamily 0909.3410}}].

\bibitem{Ji:2014eta}
Y.~Ji and A.~V. Belitsky, \emph{{Renormalization of twist-four operators in
  light-cone gauge}},
  \href{http://dx.doi.org/10.1016/j.nuclphysb.2015.03.002}{\emph{Nucl. Phys.}
  {\bfseries B894} (2015) 161--222},
  [\href{https://arxiv.org/abs/1405.2828}{{\ttfamily 1405.2828}}].

\bibitem{Braun:2008ia}
V.~M. Braun, A.~N. Manashov and J.~Rohrwild, \emph{{Baryon Operators of Higher
  Twist in QCD and Nucleon Distribution Amplitudes}},
  \href{http://dx.doi.org/10.1016/j.nuclphysb.2008.08.012}{\emph{Nucl. Phys.}
  {\bfseries B807} (2009) 89--137},
  [\href{https://arxiv.org/abs/0806.2531}{{\ttfamily 0806.2531}}].

\end{thebibliography}\endgroup

\bibliographystyle{JHEP}


\end{document}